\documentclass[10pt,jpa]{iopart} 

\usepackage{array}

\usepackage{iopams}
\expandafter\let\csname equation*\endcsname\relax

\expandafter\let\csname endequation*\endcsname\relax

\usepackage{amsmath}
\usepackage{graphicx}
\usepackage{mathrsfs}
\usepackage{amsthm}
\usepackage{amssymb}
\usepackage{xypic}
\usepackage{hyperref}
\usepackage{bm}
\usepackage{float}

\newcommand{\abs}[1]{\left| #1 \right|}
\newcommand{\norm}[2][]{\left\| #2 \right\|_{#1}}

 \newcommand{\ketbra}[3][ ]{\left| #2 \vphantom{#3} \middle \rangle _{#1} \middle \langle  #3 \vphantom{#2} \right|}

\renewcommand{\tr}[2][ ]{\mathrm{tr}_{#1}\left[#2\right]}
\newcommand{\comm}[2]{\left[ #1\vphantom{#2} , #2 \vphantom{#1} \right]}
\newcommand{\R}{\mathbb R}
\newcommand{\half}{\case{1}{2}}

\newtheorem{theorem}{Theorem}[section]
\newtheorem{lemma}[theorem]{Lemma}
\newtheorem{proposition}[theorem]{Proposition}
\newtheorem{corollary}[theorem]{Corollary}

\newcommand{\Ao}{\mathsf{A}}
\newcommand{\Bo}{\mathsf{B}}
\newcommand{\Co}{\mathsf{C}}
\newcommand{\Do}{\mathsf{D}}

\newcommand{\Jo}{\mathsf{J}}
\newcommand{\Lo}{\mathsf{L}}
\newcommand{\Mo}{\mathsf{M}}

\newcommand{\Po}{\mathsf{P}}

\newcommand{\C}{\mathbb{C}}

\newcommand{\mH}{\mathcal{H}}

\newcommand{\avec}{\bm{a}}
\newcommand{\bvec}{\bm{b}}
\newcommand{\cvec}{\bm{c}}
\newcommand{\dvec}{\bm{d}}
\newcommand{\evec}{\bm{e}}

\newcommand{\gvec}{\bm{g}}

\newcommand{\mvec}{\bm{m}}
\newcommand{\rvec}{\bm{r}}
\newcommand{\xvec}{\bm{x}}
\newcommand{\yvec}{\bm{y}}
\newcommand{\zvec}{\bm{z}}
\newcommand{\nulvec}{\bm{0}}
\newcommand{\sigvec}{\bm{\sigma}}

\newcommand{\mD}{\mathcal{D}}
\newcommand{\mDc}{\mathcal{D}_{\rm cal}}
\newcommand{\Dcal}{\Delta_{\rm cal}}

\usepackage{color}

\usepackage[normalem]{ulem}

\begin{document}

\title{Measurement uncertainty relations: characterising optimal error bounds for qubits}

{\small Topical Review, \emph{J. Phys. A: Math. Theor.} (2018), DOI: \href{https://doi.org/10.1088/1751-8121/aac729}{10.1088/1751-8121/aac729}}

\author{T.~Bullock$^{1}$, P.~Busch$^2$}
\address{$^1$ QTF Centre of Excellence, Department of Physics and Astronomy, University of Turku, Turku, FI-20014, Finland}
\address{$^2$ Department of Mathematics, University of York, York, YO10 5DD, UK}
\ead{\mailto{tjbullock89@gmail.com}, \mailto{paul.busch@york.ac.uk}}

\begin{abstract}
In standard formulations of the uncertainty principle, two fundamental features are typically cast as \emph{impossibility} statements: two noncommuting observables cannot in general both be sharply defined (for the same state), nor can they be measured jointly. The pioneers of quantum mechanics were acutely aware and puzzled by this fact, and it motivated Heisenberg to seek a mitigation, which he formulated in his seminal paper of 1927. He provided intuitive arguments to show that the values of, say, the position and momentum of a particle can at least be  \emph{unsharply} defined, and they can be measured together provided some \emph{approximation errors} are allowed. Only now, nine decades later, a working  theory of approximate joint measurements is taking shape, leading to rigorous and experimentally testable formulations of associated error tradeoff relations. Here we briefly review this new development, explaining the concepts and steps taken in the construction of optimal joint approximations of pairs of incompatible observables. As a case study, we deduce measurement uncertainty relations for qubit observables using two distinct error measures. We provide an operational interpretation of the error bounds and discuss some of the first experimental tests of such relations. 
\end{abstract}


\tableofcontents

\section{Introduction}
The year 2017 marked the 90th anniversary of Heisenberg's uncertainty principle, laid down in his fundamental paper of 1927 \cite{H27}.  Heisenberg sought to mitigate two fundamental operational limitations imposed by quantum mechanics: the impossibility of preparing  simultaneous sharp values of noncommuting quantities; and the obstruction to making arbitrarily accurate joint measurements of such observables. His aim was to understand how, in the light of these restrictions, quantum mechanics could account for the evidence of particle trajectories observed in cloud chambers.s

The degree of uncertainty in the values of an observable can be operationally quantified by some measure of the spread in the distribution of that quantity attributed by the given quantum state. The textbook \emph{preparation uncertainty relation}  thus describes the tradeoff between the standard deviations  of two incompatible observables in the same state, as obtained in accurate measurements performed separately. This is one (and the more familiar) side of the coin that is the uncertainty principle.

The other side concerns a limitation of the accuracy of \emph{approximate joint measurements} of a pair of incompatible observables. It took several decades before Heisenberg's intuitive idea, which he explained by means of semi-classical descriptions of \emph{Gedanken experiments}, could be made precise. In modern terminology, the challenge at hand is that of approximating a pair of incompatible sharp observables by two observables that are jointly measurable. It is now well understood that the price for the ability of performing such approximate joint measurements comes in the form of a \emph{measurement uncertainty relation} that describes the tradeoff between suitably defined measures of approximation errors. An account of early results and examples of measurement uncertainty relations is given in \cite{QMMT}. This book also provides a self-contained treatment of the underlying quantum theory of measurement, on which we will freely draw here. An extensive review of measurement uncertainty relations, primarily for the position and momentum of a quantum particle,  is provided in \cite{BHL2007}. The necessary conceptual developments for the formulation of these relations, including the theory of joint measurements of noncommuting unsharp observables, and the first proposals of error measures applicable in quantum mechanics, are also provided. We also draw the reader's attention to the more recent review \cite{BLWcoll}, which focuses on a critical comparison of two distinct proposals for adapting the classic Gauss root-mean-square error concept to the realm of quantum measurements.

The purpose here is to present a pedagogical review of the subject of approximate joint measurements and associated measurement uncertainty relations. We offer a systematic exposition of this subject, 
pulling together and integrating methods and results from a range of recent investigations (e.g., \cite{Ozawa2004,BH2008,Branciard2013,YuOh2014,BLW2014}). We feel that the time has come to incorporate a basic introduction of measurement uncertainty into the undergraduate quantum mechanics curriculum, and hope that the material presented here may serve as a starting point. Accordingly, we introduce relevant concepts simply for the case of discrete observables in finite-dimensional Hilbert spaces. Instead of looking, in the usual {\em ad hoc} manner, for specific forms of error tradeoff relations (e.g., as lower bounds for the product or sums of errors), we show how the problem of finding optimal joint approximations of incompatible observables is appropriately cast in terms of the associated {\em error region}, and solved by specifying its lower boundary. This programme will be illustrated for the case of qubit observables. In doing so, we will also present a number of new results, as specified below.

Following a brief introduction to relevant measurement-theoretic tools in Section \ref{sec:prelim}, we recall (Section \ref{sec:jmu}) the notions of joint measurability and joint observable in the case of observables with finitely many outcomes. In the general class of measurements represented by positive operator valued measures, the joint measurability (or compatibility) of two observables does not, in general, require their mutual commutativity, in contrast to the case of {\em sharp} observables, represented by  spectral measures (or the associated selfadjoint operators).  Among the pairs of noncommuting observables  one will find jointly measurable pairs provided these observables are sufficiently unsharp. This as yet little-noted feature of quantum mechanics (stated, for instance, in \cite{Busch86} and explained in some detail in Section \ref{sec:jmu}) thereby offers a much enlarged set of compatible observables within which to search for optimal joint approximations for a given pair of incompatible sharp observables.

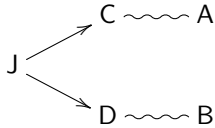
\begin{figure}[ht]
\centering
	\begin{equation*}
		\xymatrix @R=0.5pc{
			& \Co\ar@{~}[r]& \Ao\\
			\Jo\ar[ur]\ar[dr] & & \\
			& \Do\ar@{~}[r]& \Bo
		}
\end{equation*}
	\caption{\label{fig:incompapp}Two incompatible observables $\Ao$ and $\Bo$ are approximated via jointly measurable observables $\Co$ and $\Do$, with some joint observable $\Jo$. Suitably defined error measures act as figures of merit of how well $\Co,\Do$ approximate $\Ao,\Bo$, respectively.}
\end{figure}

The task of jointly approximating two incompatible observables $\Ao$ and $\Bo$ with a pair of compatible observables $\Co$ and $\Do$, respectively, (see. Fig.~\ref{fig:incompapp}) requires one to choose an appropriate measure of error, $\delta(\Ao,\Co)$, to quantify how close an approximator $\Co$ comes to matching the target observable $\Ao$. 
A variety of error measures have been proposed and explored, leading to measurement uncertainty relations of differing form.
 In Section \ref{sec:erm}, we briefly review some of these measures and introduce instances of the so-called {\em metric error}  and {\em calibration error}, both of which represent device figures of merit. These quantities meet the defining requirement for any operational error measure: that it represents a generally applicable experimental error estimation procedure, yielding an error value that can be compared with the theoretical value. As device figures of merit, such error measures are state-independent: they quantify the worst-case deviation between probability distributions of two observables, taken across all possible states. Up to now, attempts at defining {\em state-dependent} error measures have not  met the above crucial requirement of reflecting an actual, universally applicable error estimation procedure; the interpretation of  tradeoff inequalities  for such measures  as measurement uncertainty relations remains therefore problematical.
 
In Section \ref{sec:ajm} we introduce the task of identifying optimal approximations to incompatible target observables $\Ao,\Bo$ from within the set of compatible pairs of observables $\Co,\Do$. Beginning with some relevant observations, we present steps to reduce the search space to pairs $\Co,\Do$ of specific simpler types, which are still guaranteed to contain the optimisers. This follows work of \cite{BH2008}, showing that the methods apply to calibration error as well as metric error.
While the metric error region (Section \ref{sec:mer}), has been characterised mathematically by Yu and Oh in \cite{YuOh2014}, we here add a new interpretation in operational terms. We also give a new explicit derivation of the calibration error region (Section \ref{sec:cer}) as the solution of an optimisation problem. This result is closely connected with a measurement uncertainty relation due to Branciard \cite{Branciard2013} that has attracted significant attention recently; a full explanation of this remarkable coincidence is provided.
We find that different choices of error measures yield different sets of optimising approximator pairs $\Co,\Do$ (as one would expect). 
In the case of maximally incompatible target observables we also exhibit a close connection between the error region and the preparation uncertainty region. 

In Section \ref{sec:tests} we present some recent examples of experimental realisations of optimal joint approximate measurements of incompatible qubit observables. We also investigate which of these experiments constitute {\em direct tests} (suitably defined) of measurement uncertainty relations. We conclude in Section \ref{sec:discuss} with a discussion of our findings.

\section{Preliminaries}\label{sec:prelim}

We adopt the standard Hilbert space description for the case of finite-dimensional quantum systems, with $\mH$ denoting the given (complex) Hilbert space. We denote {\em vector states} by unit vectors $\psi,\xi\in\mH$  and (generally) mixed states by density operators (positive operators of unit trace) $\rho$. Vector states correspond to the pure states in the convex set of density operators, where they are represented as rank-1 projection operators $P_\psi = \ketbra{\psi}{\psi}$.

Among the set of self-adjoint linear operators acting on $\mH$, a distinguished subset is given by the set of {\em effects}; these are  the operators in the interval $ 0\leq E \leq I $, where $0,I$ are the null and identity operator, respectively, and $\le$ denotes the operator partial ordering, $A\le B$ if $\langle\psi|A\psi\rangle\le \langle\psi|B\psi\rangle$ for all $\psi\in\mH$. In other words, these operators are positive and guarantee that, for any state $\rho$, $\tr{E\rho}\le 1$.

In this paper we consider only discrete observables, represented as positive operator-valued measures (POVMs): a POVM $\Co$ in $\mH$ is defined as a map $c_i\mapsto \Co(c_i)=C_i$ from a discrete outcome set $\{c_1,\dots\,c_N\}=:\Omega_\Co$  to positive bounded operators $C_i$ (also called effects) such that $\sum_i C_i =I$. This ensures that for each state $\rho$, the map $c_i\mapsto {\rm tr}[\rho C_i]$ is a probability distribution, representing the statistics of outcomes in a measurement of $\Co$ in the state $\rho$.

Throughout, we denote POVMs by sans serif letters.  Sharp observables are described by projection-valued (or spectral) measures (PVMs): a PVM is a POVM $\Po$ whose effects $P_i = \Po(p_i)$ are projection operators ($P_i^2=P_i$) for all $i=1,2,\dots,N$. All other POVMs will be referred to as \emph{unsharp} observables. For a POVM $\Co$ with effects $\{C_1, \dots, C_N \}$, the $k^{\mathrm{th}}$ moment operator $\Co[k]$, $k\in \mathbb{N}$, is defined via
\begin{equation}
	\Co[k] = \sum_i c_i^k C_i.
\end{equation}
Note that it will usually be the case that $\Co[k]\neq \Co[1]^k$ (unless $\Co$ is sharp).

The case studies we will consider are given by so-called qubit systems, where $\mH=\C^2$. We will make use of the Bloch representation of operators acting on $\C^2$: A selfadjoint operator $A$ can be expressed in the form
\begin{equation}\label{eq:Bloch}
	A = \frac{1}{2}(a_0 I + \avec\cdot\sigvec),
\end{equation}
where $a_0\in\mathbb{R}$, $\avec\in\R^3$ is the operator's \emph{Bloch vector}, and $\sigvec= (\sigma_1,\sigma_2,\sigma_3)$ is the vector composed of the Pauli matrices,
\[
\sigma_1=\begin{pmatrix} 0&1\\1&0\end{pmatrix},\quad\sigma_2=\begin{pmatrix} 0&-i\\ i&0\end{pmatrix},\quad 
\sigma_3=\begin{pmatrix} 1&0\\ 0&-1\end{pmatrix}.
\]
 The eigenvalues of $A$ are equal to $\case{1}{2}(a_0 \pm \norm{\avec})$, and so $A$ is an effect if and only if $\norm{\avec}\leq a_0\leq2-\norm{\avec}$ (equivalently, $\norm{\avec}\le\min\{a_0, 2-a_0\}$). Furthermore, $A$ forms a projection if and only if its Bloch vector is normalised, forcing $a_0=1$.
 A density operator $\rho= \case{1}{2}(I+\rvec\cdot\sigvec)$ is characterised by $0\leq\norm{\rvec}\leq1$.
 
We will focus predominantly on binary ($\pm1$-valued) POVMs, which possess two effects. A binary qubit POVM $\Co$ has effects $\Co(\pm)=:C_\pm$ of the general form
\begin{equation}\label{eq:biased}
	C_\pm = \frac{1}{2}\big((1\pm\gamma)I \pm \cvec\cdot\sigma \big),
\end{equation}
where the positivity of the operators $C_\pm$ is equivalent to $\abs{\gamma}+\norm{\cvec}\leq 1$. 

The parameters $\gamma$ and $\norm{\cvec}(1-|\gamma|)$ are measures of \emph{bias} and \emph{sharpness} of the binary observable $\Co$. Intuitively, the effect $C_+$ may be biased towards the unit effect $I$ or the null effect $0$, depending on whether its probabilities in general tend to be closer to 1 or 0; this corresponds to a positive or negative value for $\gamma$, respectively.  The effect is \emph{unbiased}, or \emph{symmetric}, if $\gamma=0$, and we denote an unbiased observable via a ``0" subscript, e.g. $\Co_0$. Similarly, we can quantify how close this effect is to being a (nontrivial) projection, that is, a rank one projection (thereby excluding the trivial, uninformative effects $I,0$), by means of its sharpness $\norm{\cvec}(1-|\gamma|)$. This quantity is 1 if and only if $C_\pm$ are rank-1 projections, i.e., when $\gamma=0$ and $\norm{\cvec}=1$, and 0 if they are trivial, i.e., $\cvec=0$ and hence $C_\pm=\frac12(1\pm\gamma)I$. (Note that $\gamma=\pm 1$ implies $\cvec=0$). As can be seen here, unbiasedness is a necessary condition for sharpness.

These measures of bias and sharpness of an effect generalise naturally to operational measures in arbitrary dimensions (see Theorem 2 and Corollary 2 in \cite{Busch2009}): the bias of an effect $A$ can be defined as 
\[
\mathcal{B}(A)=\norm{A}-\norm{A'},
\] 
where $A':=I-A$ is the complement effect of $A$, and $\norm{\cdot}$ is the operator norm. Similarly, we can define the sharpness of $A$ as
\[
\mathcal{S}(A)=\norm{A}+\norm{A'}-\bigl(\norm{AA'}+\norm{I-AA'}\bigr).
\]
These quantities satisfy a set of natural requirements for measures of bias and sharpness, as detailed in \cite{Busch2009}.
All terms in the definitions of these measures correspond to the maximum possible probabilities of measurements of $A,A'$ or their joint measurement, represented by $AA'$ and $I-AA'$.

In what follows we will only make use of the sharpness in the case of unbiased binary qubit observables, and so $\mathcal{S}(C_\pm)=\norm{\cvec}$. This allows us to conveniently define the \emph{unsharpness} $U(\Co)$ of the observable $\Co$ via
\begin{equation}
U(\Co)^2=1-\mathcal{S}(C_\pm)^2=1-\norm{\cvec}^2.
\end{equation}
We also highlight a special subset of unbiased (or symmetric) observables $\Co$ that will be used as approximations to a sharp observable $\Ao$; these are the so-called \emph{smearings} of $\Ao$, for which the defining Bloch vector $\cvec=\norm{\cvec}\avec$. Such observables commute with $\Ao$, and the term \emph{smearing} refers to the fact that they can be defined through a procedure of mixing $\Ao$ with a trivial observable (whose effects are multiples of $I$):
\[
C_\pm=\frac12\left(I\pm\norm{\cvec}\avec\cdot\sigvec\right)=\lambda\frac12\left(I\pm\avec\cdot\sigvec\right)+(1-\lambda)\frac12 I,\quad\lambda=\norm{\cvec}.
\]

\section{Joint measurability and unsharpness}\label{sec:jmu}

For a general quantum system, two observables $\Co$ and $\Do$ with (finite) outcome sets $\Omega_\Co$ and $\Omega_\Do$, respectively, are said to be \emph{jointly measurable} if there exists an observable $\Jo$ (defined on $\Omega_\Co\times\Omega_\Do$) such that 
\begin{equation}\label{eq:compat-def}
	\Co(k)=\sum_{\ell\in\Omega_\Do}\Jo(k,\ell),\quad
	\Do(\ell)=\sum_{k\in\Omega_\Co} \Jo(k,\ell),
\end{equation} 
for any effects $\Co(k), \Do(\ell)$. Such a $\Jo$ is called a \emph{joint observable} of $\Co$ and $\Do$, and equivalently $\Co$ and $\Do$ are \emph{marginal observables} (or \emph{margins}) of $\Jo$.

\begin{figure}[ht]
\centering
	\includegraphics[width=.8\textwidth]{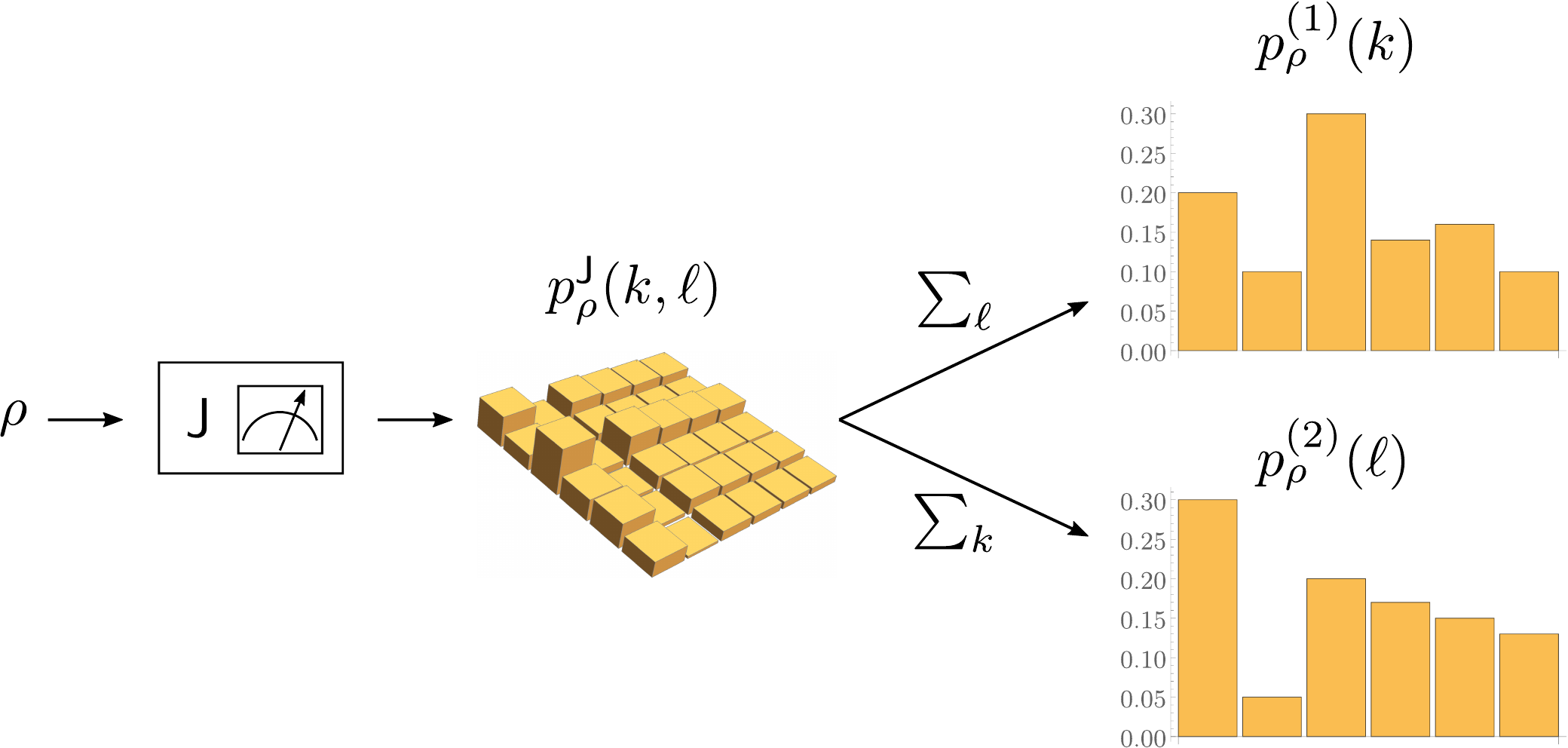}
\caption{
\label{fig:joint-meas} 
Illustration of the concept of joint measurement: systems prepared in a state described by the density operator $\rho$ enter a measuring device labelled $\Jo$, which outputs a pair of values, $(k,\ell)$. Statistics are collected over many runs, resulting in the histogram shown. The frequencies are intended to approximate the predicted quantum probabilities obtained via the Born rule as $p^\Jo_\rho(k,\ell)=\tr{\rho \Jo(k,\ell)}$, where the $\Jo(k,\ell)$ are positive operators (effects) adding to $I$. One may extract two marginal distributions by summing either over $\ell$ or $k$, giving rise to two \emph{marginal} POVMs $\Co,\Do$, respectively, which are taken to be the observables measured jointly by the scheme.}
\end{figure}

The motivation for this definition of joint measurability, which is universally adopted in the field, comes from the notion that it may be possible to devise a measurement scheme that outputs a pair of values, $(k,\ell)$, the probabilities of which are given by some observable (POVM) $\Jo$ via
\[
p^\Jo_\rho(k,\ell)=\tr{\rho \Jo(k,\ell)}.
\]
These distributions, which exist for every quantum state $\rho$ that can be used to run the experiment, give rise to two natural marginal distributions obtained by summing either over $\ell$ or over $k$:
\[
p^{(1)}_\rho(k)=\sum_\ell p^\Jo_\rho(k,\ell) \equiv \tr{\rho \Co(k)},\quad p^{(2)}_\rho(k)=\sum_k p^\Jo_\rho(k,\ell) \equiv \tr{\rho \Do(\ell)},
\]
where  the marginal observables $\Co,\Do$, as defined in \eqref{eq:compat-def}, are taken to be the observables measured jointly by the scheme. This consideration is illustrated in Fig.~\ref{fig:joint-meas}.

A sufficient condition for joint measurability of $\Co$ and $\Do$ is commutativity of their effects, i.e., $\comm{\Co(k)}{\Do(\ell)}=0$ for all $k,\ell$; but this condition is not necessary unless at least $\Co$ or $\Do$ is sharp. In that case the joint observable is unique and its effects are $\Jo(k,\ell)=\Co(k)\Do(\ell)$ \cite[Prop.~4.8]{QMMT}. 

The following result \cite{BHSS2013} illustrates the possibility of ensuring joint measurability by a procedure of smearing: two noncommuting and incompatible observables $\Ao,\Bo$ are changed by performing suitable convex combinations with so-called \emph{trivial} observables, whose effects are multiples of the identity.
\begin{proposition}\label{prop:comp-triv}
Let $\Ao:k\mapsto \Ao(k)$, $\Bo:\ell\mapsto\Bo(\ell)$ be two (possibly noncommuting and incompatible) observables with outcome spaces $\Omega_1,\Omega_2$. Then for any $\lambda\in[0,1]$ and any pair of probability distritbutions $p:k\mapsto p(k)$ on $\Omega_1$ and $q:\ell\mapsto q(\ell)$ on $\Omega_2$, the observables $\Co,\Do$ are jointly measurable, where
\begin{align}\label{eq:smear}
\Co(k)=\lambda \Ao(k)+(1-\lambda)p(k)I,\quad \Do(\ell)=(1-\lambda)\Bo(\ell)+\lambda q(\ell)I,
\end{align}
and a joint observable $\Jo$ on $\Omega_1\times\Omega_2$ is given by
\begin{equation}
\Jo(k,\ell)=\lambda \Ao(k)q(\ell)+(1-\lambda)p(k)\Bo(\ell),\quad (k,\ell)\in\Omega_1\times\Omega_2.
\end{equation}
\end{proposition}
The proof is left as an exercise for the enthusiastic reader. The observables $\Co,\Do$ are more unsharp than $\Ao,\Bo$ in an intuitive sense; this is certainly the case when $\Ao,\Bo$ are projection-valued. Moreover, if $\Ao,\Bo$ do not commute and $\lambda\not\in\{0,1\}$, then $\Co,\Do$ are noncommuting but nevertheless compatible.

It is plausible to consider $\Co,\Do$ as reasonable approximations to $\Ao,\Bo$, respectively. However, we will not dwell on making these statements more quantitative here. Instead we point out that the procedure of Proposition \ref{prop:comp-triv} is not optimal; it is easy to construct examples of compatible $\Co,\Do$ with a parameter $\mu\in[0,1]$ replacing $1-\lambda$  (and $1-\mu$ replacing $\lambda$) in the expression for $\Do$, such that $\mu>1-\lambda$, making $\Do$ sharper. This possibility is contained in the following, much stronger result for binary unbiased qubit observables.

\begin{proposition}[\cite{Busch86}]\label{prop:qc}
For two symmetric binary qubit observables $\Co,\Do$ to be jointly measurable it is necessary and sufficient that the associated Bloch vectors $\cvec,\dvec$ 
satisfy the inequality
\begin{equation}\label{compat}
	\norm{\cvec+\dvec}+\norm{\cvec-\dvec}\leq 2.
\end{equation}
\end{proposition}

It is a simple exercise to verify that this condition is equivalent to both $\norm{\cvec-\dvec}\le1-\cvec\cdot\dvec$ and $\norm{\cvec+\dvec}\le1+\cvec\cdot\dvec$.
This implies that the following defines a joint observable for $\Co,\Do$:
\begin{equation}\label{eq:jt-obs}
	\Jo_{k,\ell}=\frac{1}{4}(1+k\ell \cvec\cdot\dvec)I+\frac{1}{4}(k\cvec+\ell \dvec)\cdot\sigvec,
\end{equation}
since the positivity  of the operators  $\Jo_{k,\ell}$ is then equivalent to (\ref{compat}). 

Given that in the qubit case commutativity is equivalent to proportionality of the Bloch vectors, this example illustrates again that joint measurements of noncommuting POVMs are possible. Furthermore, it demonstrates the sufficiency of the condition (\ref{compat}) in ensuring this compatibility; for the sake of completeness, the necessity proof is given in \ref{app:necess}.

We leave the proof of the following special case of Proposition \ref{prop:qc} as an exercise.
\begin{corollary}
Two symmetric binary qubit observables $\Co,\Do$ with perpendicular Bloch vectors $\cvec,\dvec$ are jointly measurable if and only if
\begin{equation}
\norm{\cvec}^2+\norm{\dvec}^2\le 1.
\end{equation}
\end{corollary}
In particular, we may choose $\norm{\cvec}=\norm{\dvec}=1/\sqrt2$, which saturate the inequality. Then, putting $\lambda=\mu=1/\sqrt2$ and $\hat\cvec=\cvec/\norm{\cvec}$, $\hat\dvec=\dvec/\norm{\dvec}$, we have
\begin{align*}
\Co(\pm)&=\tfrac12\left(I\pm \lambda\hat\cvec\cdot\sigvec\right)=\lambda\,\tfrac12(I\pm\hat\cvec\cdot\sigvec)+(1-\lambda)\,\tfrac12 I,\\
\Do(\pm)&=\tfrac12\left(I\pm \mu\hat\dvec\cdot\sigvec\right)=\mu\,\tfrac12(I\pm\hat\dvec\cdot\sigvec)+(1-\mu)\,\tfrac12 I.
\end{align*}
Comparing with Eq.~\eqref{eq:smear}, we see that $\lambda=1/\sqrt2$ in the definition of $\Co$ whilst for $\Do$, the same value takes the place of $1-\lambda$. Clearly,
the choice for $\Do$ made in \eqref{eq:smear} is less favourable than the present choice, which gives a sharper observable that is still compatible with $\Co$.

Further analysis of the condition (\ref{compat}) shows that it is equivalent to 
\begin{equation}\label{compat-alt}
\norm{\cvec}^2+\norm{\dvec}^2\leq 1+(\cvec\cdot\dvec)^2,
\end{equation}
which can in turn be rewritten as
\begin{equation}\label{compat-u1}
\left(1-\norm{\cvec}^2\right)\left(1-\norm{\dvec}^2\right)\,\geq \, \norm{\cvec\times\dvec}^2.
\end{equation}
Recalling the measure  of unsharpness, $U(\Co)^2=1-\norm{\cvec}^2$, and observing that the right hand side can be expressed as $4\bigl\|{[C_+,D_+]}\bigr\|^2$, we obtain the following.
\begin{corollary}
Two symmetric binary qubit observables $\Co$ and $\Do$ are joint measurable if and only if they satisfy the following unsharpness tradeoff:
\begin{equation}\label{compat-u2}
U(\Co)^2U(\Do)^2\geq 4\bigl\|{[C_+,D_+]}\bigr\|^2.
\end{equation}
\end{corollary}
This inequality is a third type of uncertainty relation in addition to preparation and measurement uncertainty. It describes the degree of unsharpness the observables $\Co,\Do$ must possess in order to be jointly measurable given their degree of noncommutativity.

Surprisingly, a generalisation of Proposition \ref{prop:qc} to include biased binary qubit observables turned out much harder to prove, with several equivalent, rather complicated formulations reported in \cite{StReHe2008,BuS2010,Yu2010}.

\section{Quantifying measurement error}\label{sec:erm}

One way of comparing two observables $\Ao,\Co$    is to see how much their statistical distributions deviate from each other. This comparison makes operational sense in the context of error estimation, as it reflects the direct comparison between acquired and expected experimental data. One such distribution-comparing measure, defined in \cite{BH2008}, is the \emph{probabilistic distance} $\mD(\Co,\Ao)$ which measures the largest statistical deviation between any two effects of $\Co$ and $\Ao$ (where we assume the same outcome space, $\Omega_\Ao=\Omega_\Co$). The probabilistic distance is given by
\begin{equation}
\eqalign{
	\mD(\Co,\Ao) =2\max_{i} \sup_{\rho}\bigl|{\tr{\rho A_i}-\tr{\rho\, C_i}}\bigr| 
	= 2\max_i \norm{ A_i-C_i},
}
\end{equation}
where the supremum is taken over all density operators $\rho$. (The factor 2 is introduced for later convenience.) 
This measure forms a metric \cite{BH2008}, and we will refer to $\mD$ interchangeably as the probabilistic distance and \emph{metric error}.

For binary observables $\Ao:\pm1\mapsto A_\pm$, $\Co:\pm1\mapsto C_\pm$ the probabilistic distance reduces to
\begin{equation}
	\mD(\Co,\Ao) =2\|A_+-C_+\|.
\end{equation}
Using \Eref{eq:biased} we have the probabilistic distance $\mD(\Co,\Ao)$ between an arbitrary binary approximating observable $\Co$ and a binary sharp target observable $\Ao$ in the qubit case:
\begin{equation}\label{qubit-metric}
	\mD(\Co,\Ao) = \abs{\gamma}+\norm{\avec-\cvec}.
\end{equation}
 We also note that with respect to this measure the optimal class of approximating observables are unbiased.
 
An alternative to defining the approximation error in terms of the supremum over all states is given by the well-known procedure of calibration: the accuracy of a new measuring device is generally assessed by testing it on systems with a precise value of the target observable the device is intended to measure or estimate. This corresponds to taking the supremum not over all states but over the eigenstates of the target observable. From this we obtain the \emph{calibration error}:
\begin{equation}
\mD_{\rm cal}(\Co,\Ao)=2\max_{i} \sup_{\rho\in S(\Ao)}\bigl|{\tr{\rho A_i}-\tr{\rho\, C_i}}\bigr| , 
\end{equation}
where $S(\Ao)=\{\rho: A_i\rho=\rho \text{ for some }i\}$ is the set of eigenstates of $\Ao$.
In the case of binary qubit observables, $\Ao(\pm)=\frac12(I\pm\avec\cdot\sigvec)$, $\Co(\pm)=\frac12((1\pm\gamma)I\pm\cvec\cdot\sigvec)$, this becomes:
\begin{equation}\label{qubit-cal}
\mD_{\rm cal}(\Co,\Ao)=|\gamma|+1-\avec\cdot\cvec.
\end{equation}
Note that this still has some properties of a distance, particularly $\mD_{\rm cal}(\Co,\Ao)=0$ implying $\Co=\Ao$.  There is also monotonicity in the sense that $\mD_{\rm cal}(\Co,\Ao)$ grows with the angle between $\cvec$ and $\avec$. But $\mD_{\rm cal}$ is not a metric since the triangle inequality is not satisfied.

There are many other ways of quantifying the degree by which two observables differ, which are all suitable measures of approximation error. One example is a metric, $\Delta$, that is a quantum adaptation of the classic Gaussian root-mean-square error  based on the so-called Wasserstein-2 distance of probability measures  \cite{BLWcoll}.  In the qubit case this amounts to \cite{BLW2014}
\begin{equation}\label{eq:w2}
	\Delta(\Co,\Ao)^2= \sup_{\rho} 2\abs{-\gamma+ \rvec\cdot(\avec-\cvec)} = 2\abs{\gamma}+2\norm{\avec-\cvec}=2\mD(\Co,\Ao).
\end{equation}
An alternative to taking the supremum over all states in (\ref{eq:w2}) is again to consider just the eigenstates of $\Ao$, i.e, $\rvec=\pm \avec$, where the target probability distributions are point-like. This again gives a measure of calibration error $\Dcal$:
\begin{equation}\label{eq:w2c}
	\Dcal(\Co,\Ao)^2= 2\abs{\gamma} + 2(1-\avec\cdot\cvec)=2\mD_{\rm cal}(\Co,\Ao).
\end{equation}
As for the metric error, the optimal approximating observables for minimising the calibration error are unbiased.

Being a form of root-mean-square deviation, the metric $\Delta$ is defined so as to scale with the dimensions of the target  and approximating observables in question. This is not the case with the probabilistic distance $\mD$ or its calibration variant $\mD_{\rm cal}$ -- these quantities are defined without reference to the values of the observables; this results in their proportionality to the squared Wasserstein-2 measures in \eqref{eq:w2} and \eqref{eq:w2c}.

We mention some further alternative error measures that have been proposed and used to introduce new forms of measurement uncertainty relations. Some (but not all) of these are metric (e.g., \cite{Miyadera2008,RSH2017}). Examples of non-metric calibration error measures are given by the \emph{error bar width} \cite{BuPe2007,Miyadera2011} and an entropic measure of an average deviation of probability distributions across all eigenstates of a target observable \cite{Buscemi2014}.

The error measures defined above provide worst-case error estimates in that their values give upper bounds within the set of all states (or all eigenstates), for the deviations of the distributions of $\Ao$ and $\Co$. Such error measures  serve as figures of merit of the $\Co$ measurement as an approximation of  $\Ao$: one can be sure that the deviation between the statistics will always lie within the bounds given by this measure, whatever the state being measured. However,
the worst-case error value will be realised on different states for different target observables. Hence, if a joint measurement of $\Co,\Do$ is performed in a state that realises the worst-case $\Ao$ error, then the actual statistical deviation of $\Do$ from $\Bo$ may be smaller than the worst-case $\Bo$ error measure suggests. This observation has led to calls for state-dependent quantifications of measurement errors, but this turns out highly problematic since in generic situations it appears difficult to disentangle measurement imperfection from preparation uncertainty.

The first attempt at deducing generic measurement uncertainty relations is due to Ozawa (e.g., \cite{Ozawa2002}). It is based on a \emph{state-dependent} measure of the deviation between two observables, which is also cast as a quantum generalisation of the root-mean-square error. This quantity, called \emph{measurement noise}, gives rise to a form of measurement uncertainty relation that was claimed to constitute  a demonstration of a violation of Heisenberg's error-disturbance relation. This claim, which was brought into the limelight by experimental confirmations of Ozawa's inequality in 2012, was scrutinised and shown to be unfounded in \cite{BLWcoll}. However, as we shall show, there is a close connection between measurement noise and our calibration error  in the qubit case, which leads to a reinterpretation of the experiments testing the associated tradeoff relations for measurement noise. This will be elucidated further below.

The measurement noise quantity can be expressed as follows,
\begin{equation}\label{eq:mmt-noise}
	\varepsilon(\Co,\Ao,\rho)^2 = {\rm tr}\left[\rho(\Co[1]-\Ao[1])^2\right] + {\rm tr}\left[\rho\bigl(\Co[2]-\Co[1]^2\bigr)\right].
\end{equation}
It is immediately evident that unless the observables $\Co$ and $\Ao$ are compatible (hence unless they commute), the measurement noise cannot be determined from the measurement statistics of $\Ao$ and $\Co$ in the state $\rho$. It requires the measurement of other quantities---in the above formulation of $\varepsilon$ the operator $(\Co[1]-\Ao[1])^2$---which do not, in general, commute with $\Ao$ and $\Co$ and hence require an entirely different measurement setup that has nothing to do with the error estimation. In response to this criticism, Ozawa proposed what has become known as the three-state method \cite{Ozawa2004}. However, this still does not allow for a comparison of the theoretical quantity $\varepsilon(\Co,\Ao,\rho)$ with an error estimate obtained from the statistics of measurements of $\Ao$ and $\Co$ on the state $\rho$. Moreover, the three-state approach undermines the intended virtue of $\varepsilon$ being specific to the state $\rho$. 

The measurement noise thus does not match its intended interpretation as measurement error unless, possibly, when the approximating measurement $\Co$ is compatible with $\Ao$. If this requirement is not met, the quantity $\varepsilon$ is generally found to become unreliable as an error measure \cite{BLWcoll}.

However, it has been pointed out already by Ozawa (e.g., \cite[Eqs. 124, 125]{Ozawa2004}) that the quantity $\varepsilon(\Co,\Ao,\rho)^2$ represents an operational squared deviation if the state $\rho$ is  an eigenstate of $\Ao$. Moreover, it was noted that in such calibration situations this quantity may agree with the calibration error based on the Wasserstein-2  deviation \cite{Ozawa2013,BLWcoll}. We illustrate this connection in the qubit case.

For binary qubit observables, the measurement noise quantity assumes the following form:
\begin{equation}
\varepsilon(\Co,\Ao,\rho)^2= 2(1-\avec\cdot(\cvec+\gamma\rvec)).
\end{equation}
Interestingly, in the unbiased case, $\Co=\Co_0$, the noise measure is state-independent; in fact, we then see the connection with the calibration error (this observation was also noted in \cite{SulyokSponar2017}):
\begin{equation}\label{eq:noise-cal}
	\varepsilon(\Co_0,\Ao,\rho)^2= 2(1-\avec\cdot\cvec)=\Delta_{\rm cal}(\Co_0,\Ao)^2=2\mD_{\rm cal}(\Co_0,\Ao).
\end{equation}
Specialising further the set of approximating observables $\Co$ to those that are smearings of $\Ao$ and thus commute, $\cvec=\norm{\cvec}\avec$, we finally obtain
\begin{equation}\label{eq:D=Dcal=eps2}
\mD(\Co,\Ao)=\norm{\cvec-\avec}=1-\norm{\cvec}=1-\cvec\cdot\avec=\mDc(\Co,\Ao)=\tfrac12\varepsilon(\Co_0,\Ao,\rho)^2.
\end{equation}

\section{Approximate joint measurements: general considerations}\label{sec:ajm}

\subsection{Joint measurability}
Joint measurements of compatible observables $\Co,\Do$ can be utilised to simultaneously obtain information about a pair of incompatible sharp (target) observables $\Ao$ and $\Bo$, in the sense that the statistics  of $\Co$ and $\Do$ provide estimations of the statistics of $\Ao$ and $\Bo$, respectively. The degree of inaccuracy in these estimates will be quantified by a given choice of error measure. 
This strategy is illustrated in \Fref{fig:incompapp}.

For $\Co$, $\Do$ to be good approximations to $\Ao$, $\Bo$, respectively, it is natural to assume that the approximating observable has the same value set as the target observable. In the qubit case, $\Ao$ and $\Bo$ are necessarily binary observables, and hence we will assume that both $\Co$ and $\Do$ are also binary, so the joint observable $\Jo: (k,\ell) \mapsto \Jo(k,\ell)$, $k,\ell\in\{-1,+1\}$, satisfies 
\begin{equation*}
C_{k} = \Jo(k,+1)+\Jo(k,-1),\quad D_{\ell}=\Jo(+1,\ell)+\Jo(-1,\ell).
\end{equation*}

\subsection{Error region}

We now turn to the task of determining the optimal error bounds. This amounts to specifying the lower boundary of the \emph{error region}, that is, the set of pairs of error values that can be attained through a joint measurement of compatible  observables $\Co$ and $\Do$  approximating incompatible observables $\Ao$ and $\Bo$, respectively.\footnote{An illustration for obtaining these bounds is provided in the authors' Wolfram demonstration \emph{Optimal Joint Measurements of Qubit Observables} at \url{http://demonstrations.wolfram.com/OptimalJointMeasurementsOfQubitObservables/}.} Our aim is not merely to solve this mathematical problem but to elucidate how the optimal errors are determined as an interplay between the unsharpness of the approximators and the incompatibility of the target observables.
Before we discuss the solution for the metric and calibration errors, we make some general observations that serve to simplify the problem \cite{BH2008}.

For both errors $\mD,\mDc$, the error region is a subset of the closed square $[0,2]\times[0,2]$. Indeed, from the form of the errors, $\mD(\Co,\Ao)=|\gamma|+\norm{\avec-\cvec}$ and $\mDc(\Co,\Ao)=|\gamma|+1-\avec\cdot\cvec$, one can see that these expressions are bounded above by $|\gamma|+1+\norm{\cvec}$, and since the positivity of a general effect $C_+=\frac12(1+\gamma)I+\frac12\cvec\cdot\sigvec$ is equivalent to $|\gamma|+\norm{\cvec}\le 1$, the ultimate upper bound is 2. Moreover, for either error measure, any value above 1 can be realised by a trivial observable with effects $\lambda I,(1-\lambda) I$, which has 
$\cvec=\nulvec$ and $\lambda=\frac12(1+\gamma)\in[0,1]$. Since a trivial observable is compatible with any other observable, this means that  the complement of $[0,1)\times[0,1)$ is included in the error regions. We can therefore focus on determining the lower boundary of the error region, which will be found to be a convex curve within $[0,1]\times[0,1]$.  In the remainder of this section, $\Co,\Do$ will denote symmetric approximators. We will assume the target observables $\Ao,\Bo$ to have Bloch vectors with an angle $\theta\in(0,\pi/2]$ between them. 

The optimisation problem at hand is simplified by the fact that given our error measures, the best approximators can be found within a suitably reduced family of observables.

\subsection{Reducing the search space}\label{sec:sym}

In \cite{BH2008} it was shown that the optimal compatible approximators $\Co,\Do$ for $\pm 1$ valued qubit observables $\Ao,\Bo$ are to be found among the symmetric qubit POVMs characterised above. To be precise, we have the following result.
\begin{proposition}\label{sym-approx}
Let $\Co,\Do$ be compatible binary qubit observables used to approximate the sharp binary observables $\Ao,\Bo$. Then $\Co_0,\Do_0$, the symmetric binary observables with the same Bloch vectors $\cvec,\dvec$, are also compatible and their approximation errors with respect to $\Ao,\Bo$ are not greater:
\begin{equation}\label{better}
\eqalign{
\mD(\Co_0,\Ao)\le\mD(\Co,\Ao), \qquad
\mD_{\rm cal}(\Co_0,\Ao)\le \mD_{\rm cal}(\Co,\Ao).
}
\end{equation}
\end{proposition}
For the proof we need the following lemma, here simply formulated for the case of  observables with identical finite outcome sets.
\begin{lemma}\label{lem:convjm}
If the pairs of observables $\Co,\Do$ and $\Co',\Do'$ are compatible and $\lambda\in(0,1)$, then the pair $\Co^{(\lambda)},\Do^{(\lambda)}$ is compatible, 
where $\Co^{(\lambda)}(k)=\lambda\Co(k)+(1-\lambda)\Co'(k)$
 and $\Do^{(\lambda)}(\ell)=\lambda\Do(\ell)+(1-\lambda)\Do'(\ell)$ for all $k,\ell$.
\end{lemma}
The proof follows simply from the fact that if $\Jo,\Jo'$ are joint observables for $\Co,\Do$ and $\Co',\Do'$, respectively, then $\Jo^{(\lambda)}$ is a joint observable for  $\Co^{(\lambda)},\Do^{(\lambda)}$, where $\Jo^{(\lambda)}(k,\ell)=\lambda \Jo(k,\ell)+(1-\lambda)\Jo'(k,\ell)$.

Now, for $\Co_\pm=\frac12(1\pm\gamma)I\pm\frac 12\cvec\cdot\sigvec$, $\Do_\pm=\frac12(1\pm\delta)I\pm\frac 12\dvec\cdot\sigvec$, we define $\Co',\Do'$ via
\[
\Co'_\pm=\frac12(1\mp\gamma)I\pm\frac 12\cvec\cdot\sigvec,\qquad \Do'_\pm=\frac12(1\mp\delta)I\pm\frac 12\dvec\cdot\sigvec.
\]
Note that $\Co_0=\frac12\Co+\frac12\Co'$ and $\Do_0=\frac12\Do+\frac12\Do'$, and it follows immediately from (\ref{qubit-metric}) and (\ref{qubit-cal}) that (\ref{better}) holds. Furthermore, we claim that if $\Co,\Do$ are compatible, then so are $\Co',\Do'$, and therefore, by Lemma \ref{lem:convjm}, $\Co_0,\Do_0$ are compatible. To see this, we note that observables $\Co',\Do'$ consist of the same effects as $\tilde\Co,\tilde\Do$, where $\tilde\Co(k)=U\Co(k)U^*$, 
$\tilde\Do(\ell)=U\Do(\ell)U^*$, with $U=\evec\cdot\sigvec$ and $\evec$ is a unit vector perpendicular to both $\cvec,\dvec$. (Then $U$ is a unitary that flips 
both $\cvec\cdot\sigvec\mapsto -\cvec\cdot\sigvec$ and $\dvec\cdot\sigvec\mapsto -\dvec\cdot\sigvec$.) Now the compatibility of $\Co,\Do$ entails 
that of $\tilde\Co,\tilde\Do$ since a joint observable, $(k,\ell)\mapsto\Jo(k,\ell)$ for the former pair gives rise to a joint observable 
$(k,\ell)\mapsto U\Jo(k,\ell)U^*$ for the latter pair. We leave it as an exercise to verify that by rearrangement of the effects $U\Jo(k,\ell)U^*$ a joint observable for $\Co',\Do'$ is obtained.

We can restrict the set of compatible approximators further such that the associated Bloch vectors $\cvec,\dvec$ are in the plane spanned by $\avec,\bvec$ whilst ensuring that this smaller set still contains optimal compatible approximators.
\begin{proposition}
Let $\Ao,\Bo$ be a pair of incompatible sharp binary qubit observables, with (non-collinear) Bloch vectors $\avec,\bvec$.
Let $\Co,\Do$ be compatible symmetric binary qubit observables with Bloch vectors $\cvec,\dvec$. There are compatible symmetric binary qubit observables $\Co_0,\Do_0$ with Bloch vectors $\cvec_0,\dvec_0$ in the plane spanned by $\avec,\bvec$ such that
\begin{equation}\label{better2}
\mD(\Co_0,\Ao)\le\mD(\Co,\Ao), \qquad
\mD_{\rm cal}(\Co_0,\Ao)\le \mD_{\rm cal}(\Co,\Ao).
\end{equation}
\end{proposition}
To prove this, let $\cvec_0,\dvec_0$ be the projections of $\cvec,\dvec$ onto the plane spanned by $\avec,\bvec$. We can write $\cvec=\cvec_0+\cvec'$ and $\dvec=\dvec_0+\dvec'$. Note that $\cvec'$ and  $\dvec'$ are perpendicular to the plane spanned by $\avec,\bvec$. Given that $\Co,\Do$ are compatible, then the symmetric binary observables $\tilde\Co,\tilde\Do$ are compatible, where the associated Bloch vectors are $\tilde\cvec=\cvec_0-\cvec'$ and $\tilde \dvec=\dvec_0-\dvec'$. This follows from the fact that the compatibility inequality \eref{compat-alt} is satisfied (since $\norm{\tilde\cvec}=\norm{\cvec}$, 
$\|{\tilde\dvec}\|=\norm{\dvec}$, and $\tilde\cvec\cdot\tilde\dvec=\cvec\cdot\dvec$):
\[
\norm{\tilde\cvec}^2+\|{\tilde\dvec}\|^2\leq1+(\tilde\cvec\cdot\tilde\dvec)^2\quad\iff\quad
\norm{\cvec}^2+\norm{\dvec}^2\leq1+(\cvec\cdot\dvec)^2.
\]
Moreover, we have
\begin{equation*}
\eqalign{
\mD(\tilde\Co,\Ao)&=\norm{\cvec_0-\cvec'-\avec}=\sqrt{\norm{\cvec_0-\avec}^2+\norm{\cvec'}^2}=\mD(\Co,\Ao),\\
\mD(\tilde\Do,\Bo)&=\norm{\dvec_0-\dvec'-\bvec}=\sqrt{\norm{\dvec_0-\bvec}^2+\norm{\dvec'}^2}=\mD(\Do,\Bo),
}
\end{equation*}
and similarly
\[
\mDc(\tilde\Co,\Ao)=1-\avec\cdot\cvec_0=\mDc(\Co,\Ao),\quad \mDc(\tilde\Do,\Bo)=1-\bvec\cdot\dvec_0=\mDc(\Do,\Bo).
\]
By Lemma \ref{lem:convjm}, $\Co_0,\Do_0$ are compatible, where $\Co_0=\frac12\Co+\frac12\tilde\Co$, $\Do_0=\frac12\Do+\frac12\tilde\Do$. These are symmetric binary observables with associated Bloch vectors $\cvec_0,\dvec_0$. It is then easy to verify that
\begin{equation*}
\eqalign{
\mD(\Co_0,\Ao)&\le\mD(\Co,\Ao), \ \,\qquad \mD(\Do_0,\Bo)\le\mD(\Do,\Bo),\\
\mDc(\Co_0,\Ao)&=\mDc(\Co,\Ao), \quad \mDc(\Do_0,\Bo)=\mDc(\Do,\Bo).
}
\end{equation*}

\section{Error bounds for qubits: metric error region}\label{sec:mer}

\subsection{The optimisation problem}\label{sec:optim-metric}

The error optimisation consists of finding the smallest possible value of $\mD(\Do,\Bo)=\norm{\bvec-\dvec}$
for a given value of $\mD(\Co,\Ao)= \norm{\avec-\cvec}$, subject to the compatibility condition \eref{compat}. Thus, to be specific, one restricts the class of compatible approximators $\Co,\Do$ to those with a fixed given value of $\mD(\Co,\Ao)$, and must identify the pairs $\Co,\Do$ for which $\mD(\Do,\Bo)$ is then minimal. This mathematical problem was solved in \cite{YuOh2014}.\footnote{A more detailed presentation of the proof including steps that were not given explicitly in that paper can be found in \cite{Bullock2015}.}
Here we provide an operational interpretation of the boundary of the error region, describing it as an error tradeoff that is governed by the unsharpness of the approximating observables, their mutual noncommutativity, and the degree of incompatibility of the target observables.

The geometric significance of the constraint  \eref{compat} is readily determined:  if we fix the vector $\cvec$, say, then this inequality defines an ellipsoid within which the end points of all possible vectors $\dvec$ must lie for the corresponding observable $\Do$ to be jointly measurable with $\Co$. The semimajor axis of this ellipsoid is the diameter in the direction given by $\cvec$ in the Bloch sphere.\footnote{Note that this means that the inequality \eref{compat} entails that both vectors $\cvec,\dvec$ must have lengths not greater than 1.}
Similarly, there is an ``admissible'' ellipsoid for the vectors $\cvec$ for any fixed $\dvec$. 

For a given fixed value of $\mD(\Co,\Ao)$, take any $\cvec$ that realises this value; the best choice of $\dvec$ then is the one whose endpoint is on  the smallest circle centred at (the end point of) $\bvec$ that still intersects with the ellipsoid of allowed vectors $\dvec$ described by \eref{compat}. Hence the end point of $\dvec$ lies on the surface of the ellipsoid, and the vector $\bvec - \dvec$ is found to be normal to that surface.

\begin{figure}[t]
\centering
\includegraphics[width=.8\textwidth]{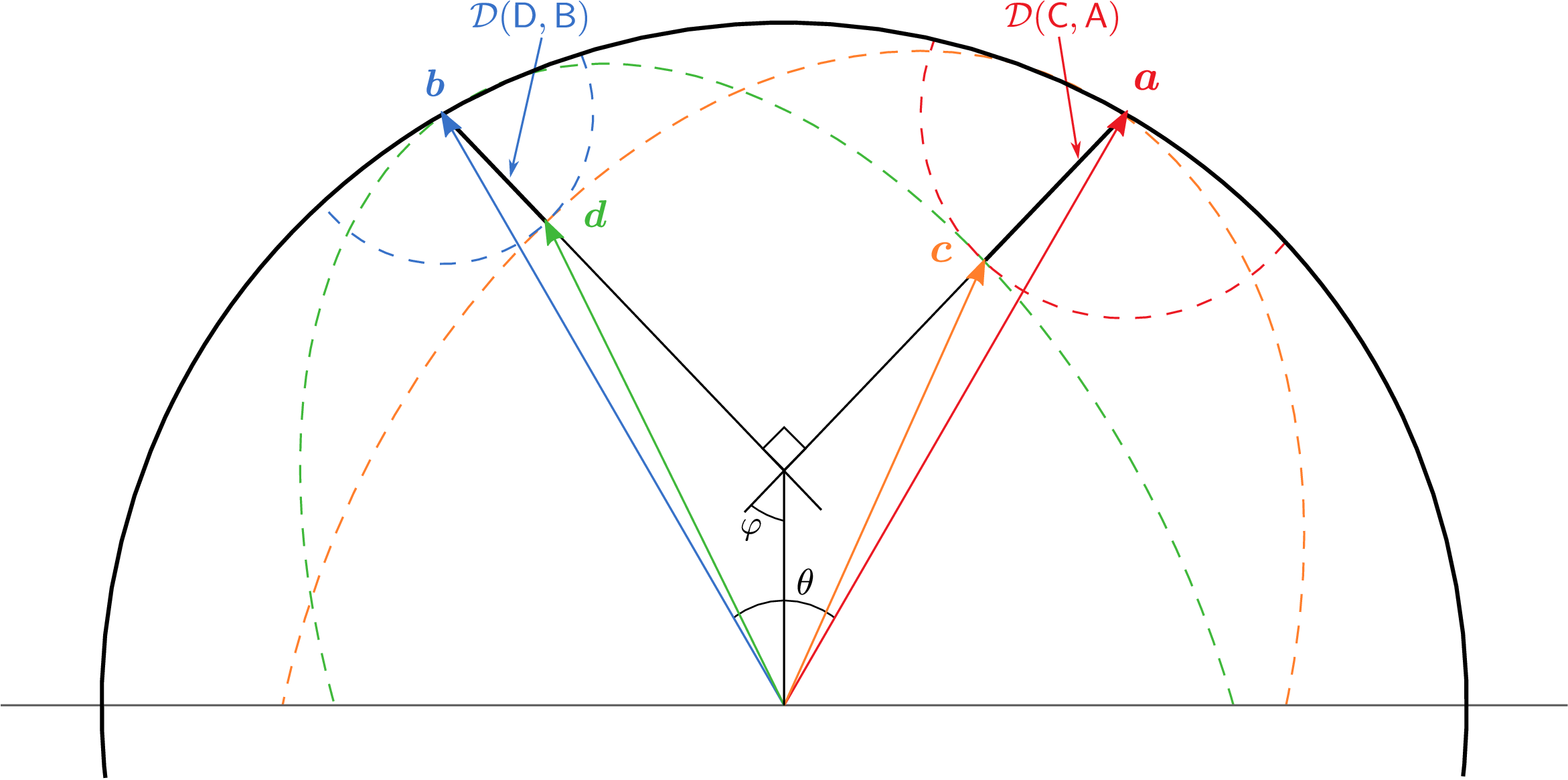}
\caption{\label{fig:YuOh-opt}Constellation of Bloch vectors $\cvec,\dvec$ giving jointly minimal metric errors.}
\end{figure}

The optimisation is complete only when $\cvec$ is also found simultaneously so that its end point lies on the surface of the corresponding ellipsoid of ``allowed" vectors $\cvec$ for the given $\dvec$, with $\avec-\cvec$ normal to that surface (\Fref{fig:YuOh-opt}).

This reasoning anticipates what is obtained by application of the method of Lagrange multipliers, where one 
needs to find the stationary points ($\nabla_{\dvec} \mathcal{L}^\Bo=0$) of the functional
\[
\mathcal{L}^\Bo(\dvec):=\norm{\bvec-\dvec}\,+\,\lambda \bigl(\norm{\cvec+\dvec}+\norm{\cvec-\dvec}-2\bigr),
\]
and similarly for $\mathcal{L}^\Ao(\cvec)$. The Euler-Lagrange equations then give
\[
\avec-\cvec\propto\frac{\cvec+\dvec}{\norm{\cvec+\dvec}}+\frac{\cvec-\dvec}{\norm{\cvec-\dvec}},\quad
\bvec-\dvec\propto\frac{\cvec+\dvec}{\norm{\cvec+\dvec}}-\frac{\cvec-\dvec}{\norm{\cvec-\dvec}}.
\]
In the case of incompatible $\Ao,\Bo$, with non-collinear vectors $\avec,\bvec$, is not hard to see from the geometry of the `admissible' ellipsoids that an optimising vector $\dvec$ cannot be collinear to $\cvec$, so that the denominators $\norm{\cvec\pm\dvec}\ne 0$. 
Since the vectors on the right-hand sides are mutually perpendicular, it follows that $\avec-\cvec$ and $\bvec-\dvec$ are perpendicular. 

There are two spheres centred at the end point of $\bvec$ with surfaces tangent to  the ellipsoid of admissible vectors $\dvec$  given a fixed $\cvec$, corresponding to a minimal and maximal distance $\norm{\bvec-\dvec}$. Similarly there will be a minimal and maximal distance $\norm{\avec-\cvec}$ within the ellipsoid of ``admissible" vectors $\cvec$. Hence there will be four stationary constellations for the optimisation problem at hand. In the following we describe the constellation of joint minima.

When  $\cvec$ and $\dvec$ have thus been found to minimise the distance   $\mD(\Do,\Bo)$ for given value of $\mD(\Co,\Ao)$ (and vice versa), then the compatibility condition \eref{compat} is saturated, giving the equation
\begin{equation}\label{eq:qucompatalt}
\norm{\cvec}^2+\norm{\dvec}^2=1+(\cvec\cdot\dvec)^2=1+M^2, \quad M=\cvec\cdot\dvec.
\end{equation} 

\subsection{Specification of metric error bounds and their operational interpretation}

Evaluation of the Euler-Lagrange equations yields the Bloch vectors associated with the optimal approximators:
\begin{equation}
\eqalign{\label{eq:BlochDopt}
	\cvec&= \frac{(\mD(\Do,\Bo)+(1-M^2)\cos\varphi)\sin\varphi\, \avec + M \mD(\Co,\Ao)\cos\varphi\, \bvec}{\sin\theta},\\
	\dvec&= \frac{(\mD(\Co,\Ao)+(1-M^2)\sin\varphi)\cos\varphi\,\bvec+M\mD(\Do,\Bo)\sin\varphi\,\avec}{\sin\theta},
}
\end{equation}
where
\begin{equation}
	\sin\varphi=\sqrt{\frac{1-\norm{\dvec}^2}{1-M^2}},\quad \cos\varphi=\sqrt{\frac{1-\norm{\cvec}^2}{1-M^2}}.
\end{equation}
As noted above, optimising $\cvec,\dvec$ cannot be collinear, so $M^2\ne 1$ and hence $\sin\varphi,\cos\varphi$ are well-defined.

The corresponding optimal (minimal) values for $\mD(\Co,\Ao)$ and $\mD(\Do,\Bo)$ are found to be 
\begin{equation}
\eqalign{\label{eq:YObound}
	\mD(\Co,\Ao)&=\frac{\sin\varphi+\sin\theta\cos\varphi}{\sqrt{1+\sin\theta\sin2\varphi}}-\sin\varphi,\\
	\mD(\Do,\Bo)&=\frac{\cos\varphi+\sin\theta\sin\varphi}{\sqrt{1+\sin\theta\sin2\varphi}}-\cos\varphi.
}
\end{equation}

We next provide an operational interpretation of this solution and the relevant quantities involved. First note that the parameter $\sin\theta$ is a measure of the incompatibility of the target observables $\Ao,\Bo$,
given by
\begin{equation}\label{eq:AB-incompat}
\sin^2\theta=\norm{\avec\times\bvec}^2=4\norm{\comm{A_+}{B_+}}^2=\tfrac14\norm{\comm{\avec\cdot\sigvec}{\bvec\cdot\sigvec}}^2.
\end{equation}
Using the unsharpness measures, $U(\Co)^2=1-\|\cvec\|^2$, $U(\Do)^2=1-\|\dvec\|^2$,  we can rewrite the compatibility equation \eref{eq:qucompatalt} in the equivalent forms
\begin{eqnarray}
1-M^2&=U(\Co)^2 + U(\Do)^2,\label{eq:MU}\\
U(\Co)^2 U(\Do)^2&=\norm{\cvec\times\dvec}^2=4\norm{[C_+,D_+]}^2.\label{eq:UCUD-com}
\end{eqnarray}

Using the above relations, we can determine $\varphi$ solely in terms of the unsharpness of the two approximating observables:
\begin{equation}\label{phi-unsharp}
\eqalign{
	\sin\varphi = \frac{U(\Do)}{\sqrt{U(\Co)^2+U(\Do)^2}},\qquad \cos\varphi = \frac{U(\Co)}{\sqrt{U(\Co)^2+U(\Do)^2}},
}
\end{equation}
and further
\begin{equation}\label{eq:varphi}
	\sin2\varphi= \frac{2U(\Co)U(\Do)}{U(\Co)^2+U(\Do)^2}.
\end{equation}
Substituting these identities in \eref{eq:YObound} we can express the bounds in terms of the unsharpness measures of $\Co$ and $\Do$ and the degree of noncommutativity $\Ao,\Bo$; we obtain:
\begin{equation}
\eqalign{\label{eq:Unsharpbound}
	\fl\qquad\mD(\Co,\Ao)&= \frac{U(\Do)+U(\Co)\sin\theta}{\sqrt{(U(\Do)+U(\Co)\sin\theta)^2+U(\Co)^2\cos^2\theta}} 
	-\frac{U(\Do)}{\sqrt{U(\Co)^2+U(\Do)^2}} ,\\
	\fl\qquad\mD(\Do,\Bo)&= \frac{U(\Co)+U(\Do)\sin\theta}{\sqrt{(U(\Co)+ U(\Do)\sin\theta)^2+U(\Do)^2\cos^2\theta}} 
	-\frac{U(\Co)}{\sqrt{U(\Co)^2+U(\Do)^2}}.
}
\end{equation}
These equations describe the desired error tradeoff, showing the dependency on the degrees of unsharpness of the approximators $\Co,\Do$ and on the degree of incompatibility of the target observables $\Ao,\Bo$.

The  quantity $M^2$ can be expressed as
\begin{equation}\label{eq:Mthetaphi}
	M^2 = \frac{\cos^2\theta}{1+\sin\theta\sin2\varphi}.
\end{equation}
Reading $M^2$ together with $\sin2\varphi$ as functions of the unsharpness measures via \eref{eq:MU}, \eref{eq:varphi}, this yields
an implicit functional relation between $U(\Co)$ and $U(\Do)$:
\begin{equation}\label{eq:unsharptrade}
\sin\theta=-\case12M^2\sin2\varphi+\sqrt{\case14M^4\sin^22\varphi+(1-M^2)}.
\end{equation}

At this point it is instructive to revisit the case $M^2=1$, for which both approximators are sharp and must therefore coincide: we now see that this entails $\sin\theta=0$, confirming that sharp approximators can only be optimal if the target observables coincide. In this case one has $\Ao=\Co=\Do=\Bo$, so that the optimal errors are both zero, as is expected.

\begin{figure}[t]
\includegraphics[width=\textwidth]{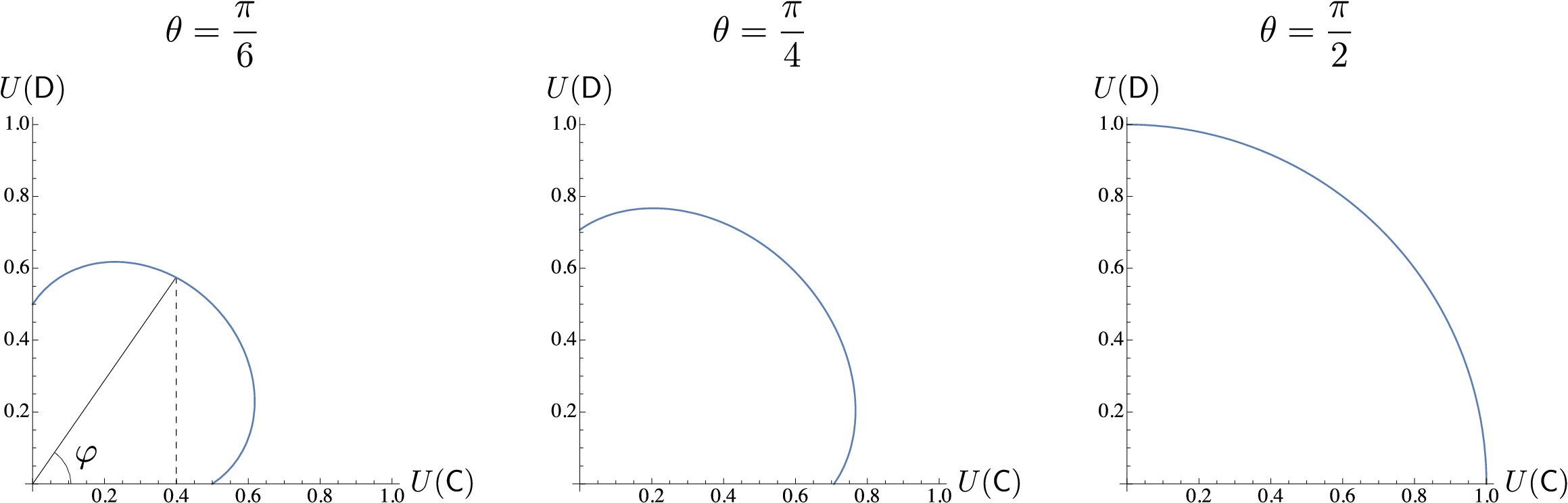}
\caption{\label{fig:Unsharp}How the tradeoff between $U(\Co)$ and $U(\Do)$ described by 
\eref{eq:unsharptrade} varies with $\theta$. The angle parameter $\varphi$ introduced in the text is also highlighted.}
\end{figure}

\Fref{fig:Unsharp} shows the unsharpness tradeoff for some target observables with different degrees of incompatibility. In general, the shape of the bound described by \eref{eq:YObound} is determined by the angle $\theta$ that quantifies the incompatibility between $\Ao$ and $\Bo$.

\subsection{Linear lower bound estimates to the metric error boundary curve}

\begin{figure}[t]
\centering
	\includegraphics[width=0.6\textwidth]{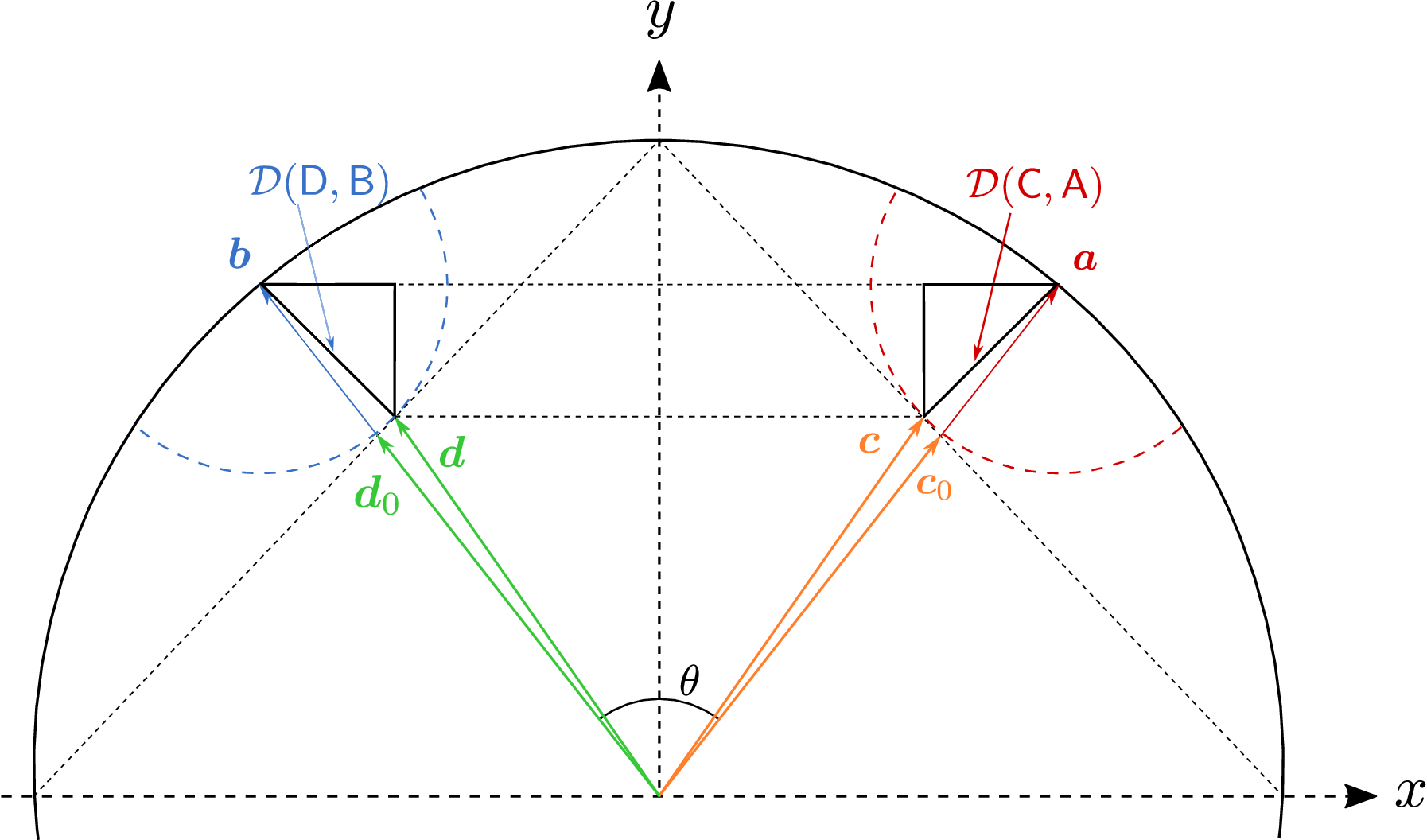}	
	\caption{\label{fig:optim-sym} Symmetric approximators yielding equal optimal error values. The optimal approximating compatible observables $\Co,\Do$ do not commute with their target observables $\Ao,\Bo$ unless the latter have perpendicular Bloch vectors $\avec,\bvec$.}
\end{figure}

We can use equations \eref{eq:Mthetaphi} and \eref{eq:YObound} to obtain a delination of the optimal error boundary curve by means of a family of tangential straight lines:
 \begin{equation}\label{eq:D-D-trade}
 \eqalign{
 \mD(\Co,\Ao)\sin\varphi+\mD(\Do,\Bo)\cos\varphi=\frac{\cos\theta}M-1.
}
 \end{equation}
 Putting $\varphi=\pi/4$ gives $\mD(\Co,\Ao)+\mD(\Do,\Bo)=\sqrt2\left[\sqrt{1+\sin\theta}-1\right]$. At this value of $\varphi$, Eq.~\eref{eq:YObound} gives 
 \begin{equation}\label{eq:sym-opt}
 \mD(\Co,\Ao)=\mD(\Do,\Bo)=\frac 1{\sqrt2}\left[\sqrt{1+\sin\theta}-1\right]
 \end{equation} 
 as an optimising error pair. This means that all points in the error region satisfy the inequality
 \begin{equation}\label{eq:lin-metric-err-bd}
 \mD(\Co,\Ao)+\mD(\Do,\Bo)\geq \sqrt2\left[\sqrt{1+\sin\theta}-1\right],
 \end{equation}
 with  equality achieved at $\mD(\Co,\Ao)=\mD(\Do,\Bo)=\left[\sqrt{1+\sin\theta}-1\right]/\sqrt2$. This inequality was first proven in \cite{BH2008}. We sketch a simple graphical proof using \Fref{fig:optim-sym}, which allows us to highlight the geometrical significance of the compatibility condition. The idea is to verify that the straight line described by equality in  \eref{eq:lin-metric-err-bd} is a tight approximation that becomes exact at the point with coordinates given in Eq.~\eref{eq:sym-opt}. The geometric construction will yield these values  and makes it evident at the same time that these are among the optimal error pairs (since the associated straight line is tangent to the convex boundary curve at this point).

 In the figure we consider sharp target and unbiased approximating observables whose Bloch vector pairs $\avec,\cvec$  and $\bvec,\dvec$ are arranged mirror symmetrically with respect to the vertical line through the origin. Recall that $\avec,\bvec$ are unit vectors. Choosing $\cvec,\dvec$ of equal length ensures that $\mD(\Co,\Ao)=\mD(\Do,\Bo)$. Optimal compatible approximators will satisfy $\norm{\cvec+\dvec}+\norm{\cvec-\dvec}=2$. Since $\norm\cvec=\norm\dvec$, the vectors $\cvec+\dvec$ and $\cvec-\dvec$ are vertically and horizontally oriented, hence perpendicular. This entails that the endpoints of $\cvec$ and $\dvec$ lie on the lines $x+y=1$ and
$y-x=1$, respectively. The circles around the endpoints of  $\avec,\bvec$ that are respectively tangent to these lines define the points (vectors $\cvec,\dvec$) of best approximation (smallest error). The corresponding vectors $\cvec,\dvec$ are thus determined by the condition that $\avec-\cvec$ is perpendicular to
 the line $x+y=1$ and $\bvec-\dvec$ is perpendicular to the line $y-x=1$. (Incidentally this shows that the optimal approximators are not, in general, smeared versions of the target observables: $\cvec,\dvec$ are collinear to $\avec,\bvec$, respectively, only when $\avec,\bvec$ are perpendicular.)
 
One can now easily work out that the optimal error values in this case are those given in Eq. \eref{eq:sym-opt}. The horizontal and vertical sides of the black equilateral triangles are 
 \[
 \tfrac12\bigl[\norm{\avec+\bvec}-\norm{\cvec+\dvec}\bigr]=\tfrac12\bigl[\norm{\avec-\bvec}-\norm{\cvec-\dvec}\bigr]=\tfrac1{\sqrt2}\mD(\Co,\Ao)=\tfrac1{\sqrt2}\mD(\Do,\Bo).
 \]
 This gives:
 \[
 \eqalign{
 \mD(\Co,\Ao)=\mD(\Do,\Bo)&=\tfrac{\sqrt2}4\bigl[\norm{\avec+\bvec}-\norm{\cvec+\dvec}\bigr]+\tfrac{\sqrt2}4\bigl[\norm{\avec-\bvec}-\norm{\cvec-\dvec}\bigr]\\
 &=\tfrac{\sqrt2}4\bigl[ \norm{\avec+\bvec}+\norm{\avec-\bvec} - \norm{\cvec+\dvec}-\norm{\cvec-\dvec}\bigr] \\
 &=\tfrac{\sqrt2}2\bigl[\cos\tfrac\theta2+\sin\tfrac\theta2-1\bigr]\\
 &=\tfrac{\sqrt2}2\bigl[\sqrt{1+\sin\theta}-1\bigr].
 }
 \]
 Here we used the relations
 $ \norm{\avec-\bvec}=2\sin\tfrac\theta2$,  $\norm{\avec+\bvec}=2\cos\tfrac\theta2$, the compatibility $\norm{\cvec+\dvec}+\norm{\cvec-\dvec}=2$,
 and the identity $\sin(\theta/2)+\cos(\theta/2)=\sqrt{1+\sin\theta}$.
 
It is interesting to observe that the optimal approximators $\Co,\Do$ in this case are not smearings of $\Ao,\Bo$ and therefore do not commute with their target observables.
 
\subsection{Circular outer bound for the metric error region} 
 
\begin{figure}[t]
\centering
	\includegraphics[width=1\textwidth]{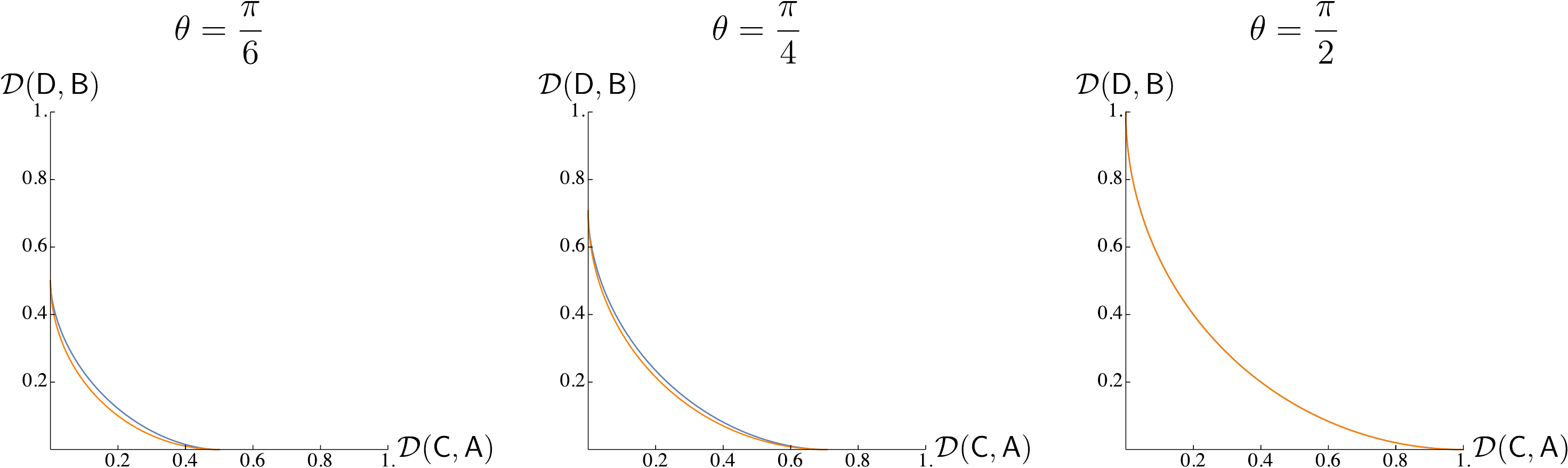}	
	\caption{\label{fig:metric-error-region}Boundary of metric error region (blue) together with (outer) circular bound (orange) for different degrees of incompatibility 
	($\theta=\frac\pi6,\frac\pi4,\frac\pi2$). In the maximally incompatible case, the two curves coincide.}
\end{figure}

Finally we remark that the exact convex boundary curve lies slightly above a circular segment of radius $\sin\theta$ centred at $(\sin\theta,\sin\theta)$. Thus, any point in the  error region with $\mD(\Co,\Ao),\mD(\Do,\Bo)\leq \sin\theta$ satisfies
\begin{equation}\label{eq:circular-approx-metric-bd}
\bigl( \mD(\Co,\Ao)-\sin\theta\bigr)^2+\bigl(\mD(\Do,\Bo)-\sin\theta\bigr)^2\leq \sin^2\theta,
\end{equation}
with equality achievable at the endpoints of the circular boundary segment, $\bigl(\mD(\Co,\Ao),\mD(\Do,\Bo)\bigr)=(\sin\theta,0)$ (at $\varphi=0$) and $\bigl(\mD(\Co,\Ao),\mD(\Do,\Bo)\bigr)=(0,\sin\theta)$ (at $\varphi=\pi/2$). \Fref{fig:metric-error-region} shows the shape of the exact boundary of the error region and the circular outer estimate for different values of $\theta$. Inequality \eref{eq:circular-approx-metric-bd} shows how the error pairs are bounded away from the origin by the measure of incompatibility of the target observables $\Ao,\Bo$,
namely, the quantity $\sin\theta=\frac12\bigl\|{[\avec\cdot\sigvec,\bvec\cdot\sigvec]}\bigr\|$.

\subsection{Minimal metric errors for maximally incompatible qubit observables}

As an important special case we consider the error tradeoff for maximally incompatible target observables, for which $\theta=\frac\pi2$.  In this case we have $M^2=0$, hence the compatibility condition becomes $U(\Co)^2+U(\Do)^2=1$, that is, $\norm{\cvec}^2+\norm{\dvec}^2=1$, with $\cvec,\dvec$ perpendicular. Eq. \eref{eq:BlochDopt} gives:
\begin{equation}
	\cvec =\sin\varphi\, \avec=U(\Do)\avec, \quad \dvec =\cos\varphi\, \bvec=U(\Co)\bvec.
\end{equation}
 The optimal approximators $\Co$ and $\Do$ are thus smeared versions of the sharp target observables $\Ao$ and $\Bo$, respectively, with associated errors given by 
 Eq.~\eref{eq:YObound},
\begin{equation}\label{eq:Dmaxincomp}
	\mD(\Co,\Ao)= 1- \sin\varphi, \qquad \mD(\Do,\Bo)=1-\cos\varphi,
\end{equation}
For the domain that we are considering, $\varphi\in[0,\pi/2]$, this defines the lower left quadrant of a unit circle centred at $(1,1)$; hence the full error region is given by all value pairs
$\bigl(\mD(\Co,\Do),\mD(\Do,\Bo)\bigr)$ with either $\mD(\Co,\Ao)\ge 1$ or $\mD(\Do,\Bo)\ge 1$, or else, when both $\mD(\Co,\Ao)\le 1$, $\mD(\Do,\Bo)\le 1$, then
\begin{equation}\label{eq:max}
(\mD(\Co,\Ao)-1)^2+(\mD(\Do,\Bo)-1)^2\leq1.
\end{equation}
The limiting case of equality defining the  exact lower boundary curve.
 
We are now in a position to observe a remarkable concurrence of preparation uncertainty, measurement uncertainty, and the unsharpness required by compatibility in the case of a maximally incompatible pair of sharp qubit observables $\Ao,\Bo$. First we note a simple form of preparation uncertainty relation for such a pair.

The variances of $\Ao,\Bo$ are $\Delta(\Ao,\rho)^2=1-(\avec\cdot\rvec)^2$, $\Delta(\Bo,\rho)^2=1-(\bvec\cdot\rvec)^2$; then,  since $\avec,\bvec$ are perpendicular, we have 
$(\avec\cdot\rvec)^2+(\bvec\cdot\rvec)^2\le\norm{\rvec}^2$, and so 
\begin{equation}\label{eq:2-pur}
\Delta(\Ao,\rho)^2+\Delta(\Bo,\rho)^2\ge2-\norm{\rvec}^2\ge 1.
\end{equation}
This determines the \emph{(preparation) uncertainty region} of $\Ao,\Bo$ as the subset of $[0,1]\times[0,1]$ that is the complement of the intersection with the open unit ball centred at (0,0):
\begin{equation}
{\rm PUR}_\Delta(\Ao,\Bo)=\left\{(\Delta \Ao,\Delta\Bo)\,:\, \Delta\Ao^2+\Delta\Bo^2\ge 1\right\}\,\cap\,\bigl([0,1]\times[0,1]\bigr).
\end{equation}
 We then have the following.
\begin{theorem}\label{thm:pum}
Let $\Ao,\Bo$ be a pair of maximally incompatible sharp observables with perpendicular Bloch vectors $\avec,\bvec$. Let $\Co,\Do$ be a pair of compatible symmetric binary observables with Bloch vectors $\cvec,\dvec$ aligned with $\avec,\bvec$, respectively, $\cvec=\norm\cvec\avec,\dvec=\norm\dvec\bvec$. The following statements are equivalent.
\begin{enumerate}
\item[(a)] $\Co,\Do$ are jointly measurable.
\item[(b)] $\Co,\Do$ satisfy the \emph{unsharpness tradeoff}:
\begin{equation}
\norm\cvec^2+\norm\dvec^2\le 1\quad \text{that is: }\quad U(\Co)^2+U(\Do)^2\ge 1.
\end{equation}
\item[(c)] $\Jo$ given in \eref{eq:jt-obs} is a joint observable for $\Co,\Do$, and it assumes the form 
\begin{equation}\label{eq:joint-state}
J_{k\ell}=\frac12\rho_{k\ell}=\frac14\bigl(I+(k\cvec+\ell\dvec)\cdot\sigvec\bigr).
\end{equation}
\item[(d)] $\Co,\Do$ satisfy the \emph{measurement uncertainty tradeoff}: 
\begin{equation}
(\mD(\Co,\Ao)-1)^2+(\mD(\Do,\Bo)-1)^2\leq1.
\end{equation}
\item[(e)] The operators $\rho_{k\ell}$ in Eq.~\eref{eq:joint-state} are positive and of trace 1 and hence satisfy the \emph{preparation uncertainty tradeoff}:
\begin{equation}
\Delta(\Ao,\rho_{k\ell})^2+\Delta(\Bo,\rho_{k\ell})^2\ge 1.
\end{equation} 
\end{enumerate}
\end{theorem}  
 These equivalences are due to the fact that $\Delta(\Ao,\rho_{k\ell})^2=1-\norm\cvec^2$, $\Delta(\Bo,\rho_{k\ell})^2=1-\norm\dvec^2$, and $\mD(\Co,\Ao)=\norm{\cvec-\avec}=1-\norm\cvec$, $\mD(\Do,\Bo)=\norm{\dvec-\bvec}=1-\norm\dvec$. They show that optimal approximations are achieved exactly at the onset of compatibility and precisely when the state operators generating the associated joint observable have minimal preparation uncertainty for the target observables. This connection was found to be rooted in the $\mathbb{Z}_2\times\mathbb{Z}_2$-covariance of the joint observable $\Jo$ \cite{BLW2014,Werner2016}.

 It is possible to express the correspondence between preparation uncertainty and measurement uncertainty emerging in the above theorem even more explicitly. To this end we note that we may define a measure of uncertainty that is more directly adapted to the metric error measure. We put
 \begin{equation*}
 \delta(\Ao,\rho)=1-|\avec\cdot\rvec|=1-\sqrt{1-\Delta(\Ao,\rho)^2},\quad \delta(\Bo,\rho)=1-|\bvec\cdot\rvec|=1-\sqrt{1-\Delta(\Bo,\rho)^2}.
 \end{equation*}
 For the states $\rho_{k\ell}$ and $\cvec=\norm{\cvec}\avec$, $\dvec=\norm{\dvec}\bvec$, this gives:
 \[
 \delta(\Ao,\rho_{k\ell})=1-\avec\cdot\cvec=1-\norm{\cvec}=\mD(\Co,\Ao),\quad \delta(\Bo,\rho_{k\ell})=1-\bvec\cdot\dvec=1-\norm{\dvec}=\mD(\Do,\Bo).
 \]
 \begin{corollary}\label{cor:MUR=PUR}
 Let $\Co,\Do$ be a compatible pair of symmetric binary observables as characterised in Theorem \ref{thm:pum}. Then the observable $\Jo$ and the states $\rho_{k\ell}$ given there trace out the nontrivial part of the metric error region and the full uncertainty region for $\Ao,\Bo$, respectively, and we have the following (easily verifiable) result.
 \begin{equation}
 {\rm MUR}_\mD(\Ao,\Bo)\cap\bigl([0,1]\times[0,1]\bigr)={\rm PUR}_\delta(\Ao,\Bo),
 \end{equation}
 where
 \begin{align*}
 {\rm MUR}_\mD(\Ao,\Bo)&=\left\{\bigl(\mD(\Co,\Ao),\mD(\Do,\Bo)\bigr)\,:\,\Co,\Do \text{ are compatible binary observables}\right\},\\
 {\rm PUR}_\delta(\Ao,\Bo)&=\left\{\bigl(\delta(\Ao,\rho),\delta(\Bo,\rho)\bigr)\,:\, \rho \text{ is any state}\right\}.
 \end{align*}
 \end{corollary}

 \subsection{Summary: the incompatibility-unsharpness-error tradeoff for metric error}
 
 In \cite{YuOh2014}, a thorough discussion of the boundary curve of the admissible region of points $\bigl(\mD(\Co,\Ao),\mD(\Do,\Bo)\bigr)$ is given with $\varphi$ as the parameter, including, in particular, its characterisation via the family  of equations \eref{eq:D-D-trade}. Here we have added the interpretation in terms of the degrees of unsharpness of $\Co$ and $\Do$, and established a functional relation between these parameters. We have seen that the optimal approximations strike a balance between the necessary degrees of unsharpness required to ensure compatibility and the magnitude of the approximation errors, and this balance is governed by the degree of incompatibility of the target observables. It becomes evident that a certain degree of noncommutativity of $\Co,\Do$ is usually favourable for improving the approximations, but too much noncommutativity requires more unsharpness (for compatibility), which drives the errors up again.

\section{Error bounds for qubits: calibration error region}\label{sec:cer}

We now turn to the similar task of determining the lower boundary curve of the admissible region of the calibration error pairs, $\bigl(\mDc(\Co,\Ao),\mDc(\Do,\Bo)\bigr)$, under the constraint of compatibility for the approximators $\Co,\Do$. As before, the optimising pairs of compatible approximators can be chosen to be symmetric binary observables.

Thus, fixing an error value $\mDc(\Co,\Ao)=1-\avec\cdot\cvec$ and an associated Bloch vector $\cvec$, we wish to determine the smallest $\mDc(\Do,\Bo)=1-\bvec\cdot\dvec$ subject to $\cvec$ and $\dvec$ satisfying inequality \eref{compat}. Optimality is again  to be obtained by vectors $\cvec,\dvec$ pointing to the surfaces of their respective compatibility ellipses (in the plane spanned by $\avec,\bvec$), thereby saturating  \eref{compat}.

Using the Lagrange multiplier method, we need to find the stationary points ($\nabla_{\dvec} \tilde{\mathcal{L}}^\Bo=0$) of the functional
\[
\tilde{\mathcal{L}}^\Bo(\dvec):=1-\bvec\cdot\dvec\,+\,\lambda \bigl(\norm{\cvec+\dvec}+\norm{\cvec-\dvec}-2\bigr),
\]
and similarly for $\tilde{\mathcal{L}}^\Ao(\cvec)$,  the Euler-Lagrange equations give
\begin{equation}\label{eq:EL2}
\avec\propto \frac{\cvec+\dvec}{\norm{\cvec+\dvec}}+\frac{\cvec-\dvec}{\norm{\cvec-\dvec}},\quad
\bvec\propto\frac{\cvec+\dvec}{\norm{\cvec+\dvec}}-\frac{\cvec-\dvec}{\norm{\cvec-\dvec}}.
\end{equation}
This is well-defined if $\dvec\neq\pm\cvec$. Then, since the vectors on the right-hand sides are mutually perpendicular, it follows that a local extremum can only exist in the case of orthogonal vectors $\avec,\bvec$, that is, maximally incompatible target observables $\Ao,\Bo$. Otherwise, if $\avec\not\perp\bvec$, the optimiser pairs, represented by Bloch vector pairs $(\cvec,\dvec)$, must lie on the boundary of the Bloch sphere, so that compatibility requires $\dvec=\pm\cvec$ (since $\Co,\Do$ are sharp). We treat both cases separately.

\subsection{Minimal calibration errors for maximally incompatible qubit observables}

In this case, $\avec\cdot\bvec=0$, Eq.~\eref{eq:EL2} entails that the optimising Bloch vectors $\cvec,\dvec$ (with $\dvec\neq\pm\cvec$) lie in the interior of the equatorial circle containing $\avec,\bvec$. Evaluating the Euler-Lagrange equations we obtain (noting that $M^2<1$)
\begin{equation}\label{eq:minmaxincomp}
\eqalign{
	\avec = \frac{\cvec - M \dvec}{(1-M^2)\sin\varphi},\qquad	\bvec = \frac{\dvec - M \cvec}{(1-M^2)\cos\varphi},
}
\end{equation}
with $M$ and $\varphi$ of the form given above. In view of \eref{eq:qucompatalt}, the vectors of the right-hand sides are easily confirmed to be mutually orthogonal.
The  Bloch vectors for the optimising approximators $\Co,\Do$ are 
\begin{equation}\label{eq:opt-cal-vec-orth}
\cvec=\sin\varphi\,\avec+M\cos\varphi\,\bvec,\quad \dvec=M\sin\varphi\,\avec+\cos\varphi\,\bvec,
\end{equation}
yielding the optimal calibration error values,
\begin{equation}\label{eq:opt-cal-err-orth}
\mDc(\Co,\Ao)=1-\avec\cdot\cvec=1-\sin\varphi,\quad \mDc(\Co,\Ao)=1-\bvec\cdot\dvec=1-\cos\varphi.
\end{equation}
The compatibility condition \eref{compat-alt}, $\norm\cvec^2+\norm\dvec^2=1+M^2$, is readily verified. It can be written as a tradeoff for the unsharpness parameters:
\begin{equation}\label{eq:opt-cal-u}
U(\Co)^2+U(\Do)^2=1-M^2.
\end{equation}
While we originally excluded the case $M^2=1$, the optimal error values do not depend on $M$ and the form of the solution \eref{eq:opt-cal-vec-orth} does allow for $M=\pm1$ to be included.
This gives two pairs of extremal (unit vetor) solutions, $\cvec_1=\dvec_1=\mvec$ and $\cvec_{-1}=-\dvec_{-1}$, and it is found that there is no constraint on the unsharpness of the optimal approximators as the bound in \eref{eq:opt-cal-u} is 0.

The parameter $\varphi$ can be eliminated, giving an explicit relation between the errors as a description of the boundary of the calibration error region:
\begin{equation}\label{eq:cal-bd}
\bigl(\mDc(\Co,\Ao)-1\bigr)^2+\bigl(\mDc(\Do,\Bo)-1\bigr)^2=1.
\end{equation}
The boundary is thus given as the segment of the unit circle with centre (1,1) within the square $[0,1]\times[0,1]$. We summarise:
\begin{proposition}\label{prop:cal-mur-orth}
The calibration error region for a pair of maximally incompatible sharp qubit observables $\Ao,\Bo$ is fully described by a single measurement uncertainty relation, given by the conditional inequality
\begin{equation}\label{eq:cal-mur-orth}
\bigl(\mDc(\Co,\Ao)-1\bigr)^2+\bigl(\mDc(\Do,\Bo)-1\bigr)^2\le 1 \quad\text{if}\ \mDc(\Co,\Ao)\leq 1,\ \mDc(\Do,\Bo)\leq 1.
\end{equation}
\end{proposition}
The observant reader will immediately notice that  inequalities \eref{eq:cal-mur-orth} and \eref{eq:max} are indeed identical and hence the metric and calibration error regions actually coincide in the case of maximally incompatible target observables ($\avec\cdot\bvec=0$). This is most easily understood by noting that for both error measures, the sets of optimising approximations contain the observables $\Co,\Do$ that are smearings of $\Ao,\Bo$, that is, $\cvec=\norm{\cvec}\avec$, $\dvec=\norm{\dvec}\bvec$, with $\norm{\cvec}^2+\norm{\dvec}^2=1$. For these approximations, the metric and calibration errors coincide, as noted in Eq.~\eref{eq:D=Dcal=eps2}.

\begin{figure}[t]
\centering
	\includegraphics[width=0.6\textwidth]{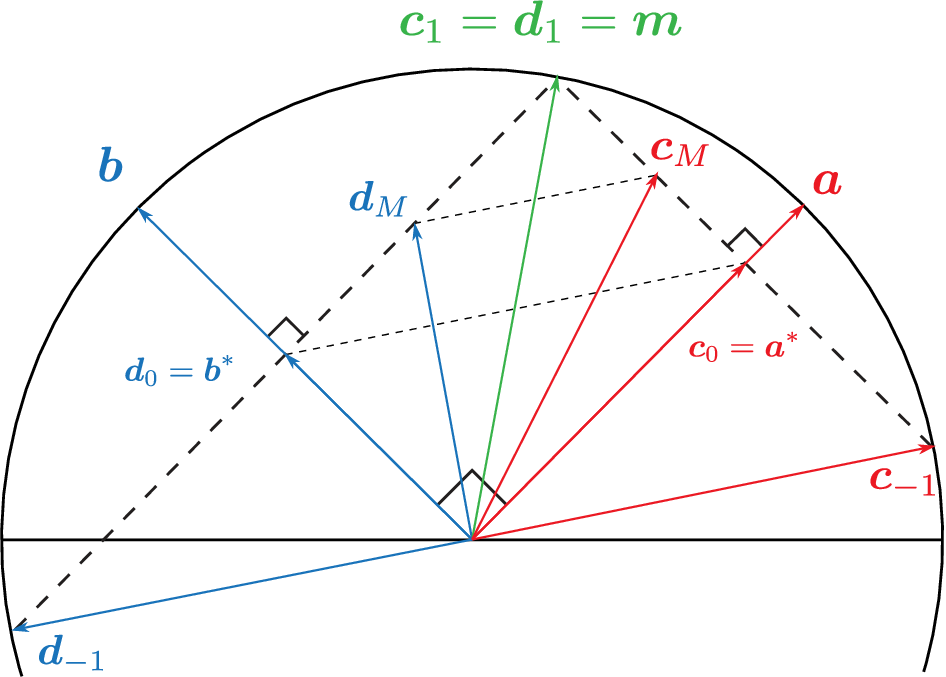}	
	\caption{\label{fig:perp}Pairs of compatible observables $\Co_M,\Do_M$ that saturate the bound given in \eref{eq:cal-bound} in the case $\avec\perp\bvec$. The associated vectors $\cvec_M$ and $\dvec_M$ are related such that $\cvec_M-\dvec_M$ and  $\cvec_0-\dvec_0$ are collinear. This includes the pair $\cvec_{-1},\dvec_{-1}$.}
\end{figure}%

It is also interesting to visualise the variety of compatible pairs $\Co,\Do$ realising one and the same point $\bigl(\mDc(\Co,\Ao),\mDc(\Do,\Bo)\big)$ on the lower boundary of that region, hence satisfying Eq.~\eref{eq:cal-bd} (Figure \ref{fig:perp}). 
Put $\mvec=\avec^*+\bvec^*$, where 
\begin{equation*}
\avec^*=(\avec\cdot\mvec)\avec=\sin\varphi\,\avec.\quad \bvec^*=(\bvec\cdot\mvec)\bvec=\cos\varphi\,\bvec.
\end{equation*}
Then we observe that both pairs $\Co_0,\Do_0$, with $\cvec_0=\avec^*$, $\dvec_0=\bvec^*$, and $\Co_1,\Do_1$, with $\cvec_1=\dvec_1=\mvec$, are compatible pairs realising the optimal error values \eref{eq:opt-cal-err-orth}. In general, the family of vector pairs $\cvec_M,\dvec_M$ of \eref{eq:opt-cal-vec-orth}, with $M\in[-1,1]$ constitute compatible approximating observables $\Co_M,\Do_M$ with the same optimal error values; these vectors can be written as convex combinations of $\avec^*+\bvec^*,\pm(\avec^*-\bvec^*)$,
\begin{equation}\label{eq:all-cal}
\begin{aligned}
\begin{array}{r@{\;}>{{}\displaystyle}l}
\cvec_M&=\tfrac12(1+M)(\avec^*+\bvec^*)+\tfrac12(1-M)(\avec^*-\bvec^*),\\[6pt]
\dvec_M&=\tfrac12(1+M)(\avec^*+\bvec^*)+\tfrac12(1-M)(\bvec^*-\avec^*),
\end{array}\quad -1\leq M\leq 1.
\end{aligned}
\end{equation}
Thus, remarkably, the calibration error selects an infinity of compatible approximating pairs of observables for each error pair on the optimal boundary of the error region when the target observables are maximally incompatible. Among these it allows, according to Eq.~\eref{eq:opt-cal-u}, an optimal approximator pair with vanishing unsharpness, $\cvec_1=\dvec_1=\mvec$.

\subsection{Minimal calibration errors for non-maximally incompatible qubit observables}\label{sec:cal}

\begin{figure}[t!]
\centering
	\includegraphics[width=.6\textwidth]{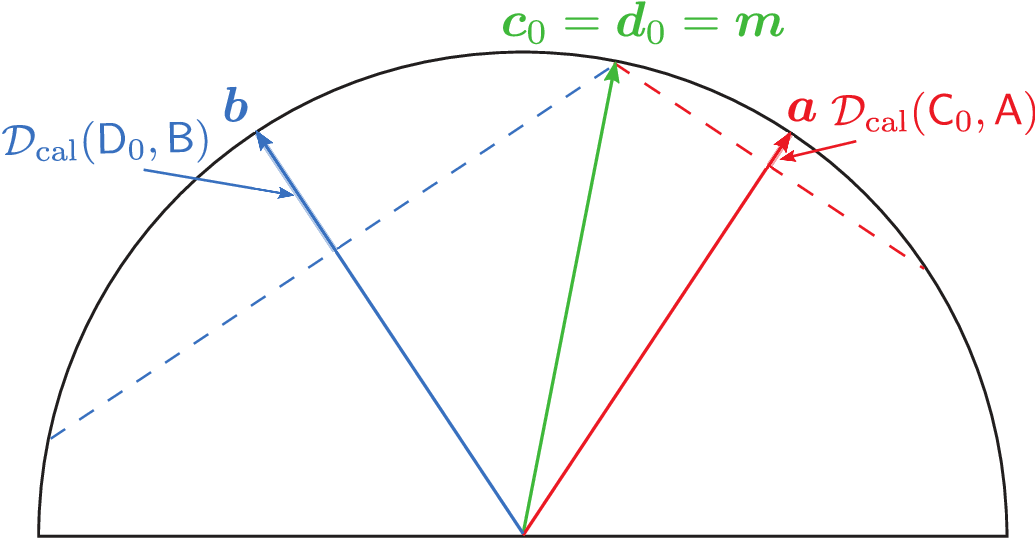}	
	\caption{\label{fig:calmin} Realising  minimal calibration error $\mDc(\Do,\Bo)$ for fixed $\mDc(\Co,\Ao)$ in the case $\avec\not\perp\bvec$. 
	The smallest possible value for $\mDc(\Do,\Bo)$ occurs when $\dvec_0=\mvec$, giving $\Do_0$, 
	in which case the unique choice $\cvec_0=\mvec$ gives an observable $\Co_0$ that is trivially compatible with $\Do_0=\Co_0$.}
\end{figure}

In the case of non-maximally incompatible target observables, $\avec\cdot\bvec\neq 0$, the optimisers must have Bloch vectors on the boundary of the Bloch sphere.
In order to see this, consider \Fref{fig:calmin}. Fix a value $\mDc(\Co,\Ao)$, with some $\cvec$ pointing to the line segment perpendicular to $\avec$ that contains the endpoint of $\avec(\avec\cdot\cvec)=\avec\bigl(1-\mDc(\Co,\Ao)\bigr)$. The ellipse of admissible vectors $\dvec$ has a tangent perpendicular to $\bvec$, thus fixing the smallest error value to be 
$\mDc(\Do,\Bo)=1-\bvec\cdot\mvec=\norm{\bvec-\bvec^*}$ such that $\Co,\Do$ are compatible for the given $\cvec$. This minimum value of $\mDc(\Do,\Bo)$ given the value 
$\mDc(\Co,\Ao)=1-\avec\cdot\mvec=\norm{\avec-\avec^*}$ is reached where the line segment perpendicular to $\bvec$ inside the Bloch sphere hits the endpoint, $\mvec$, nearest to $\bvec$ of the given line segment perpendicular to $\avec$ inside the Bloch sphere. The associated vectors $\dvec_0=\cvec_0=\mvec$ realise commuting sharp observables, $\Do_0=\Co_0$. This is the unique solution to the optimisation problem at hand if $\avec,\bvec$ are not perpendicular.

\begin{figure}[t!]
\centering
	\includegraphics[width=\textwidth]{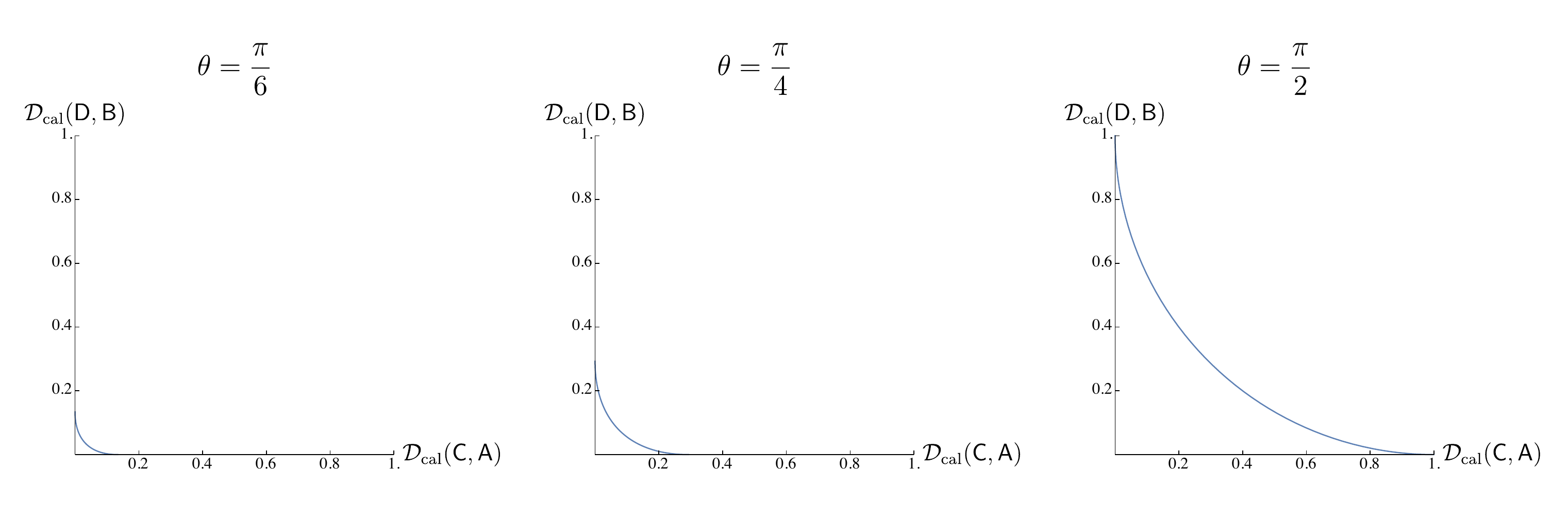}	
	\caption{\label{fig:calboundary} Lower boundary of the calibration error region for different degrees of incompatibility ($\theta=\frac\pi6,\frac\pi4,\frac\pi2$).}
\end{figure}

The calibration error values are
\begin{equation}\label{eq:opt-cal-err}
\eqalign{
\mDc(\Co,\Ao)&=1-\avec\cdot\mvec=\norm{\avec-\avec^*}=1-\cos\phi,\\
\mDc(\Do,\Bo)&=1-\bvec\cdot\mvec=\norm{\bvec-\bvec^*}=1-\cos(\theta-\phi),
}
\end{equation}
where $\phi$ is the angle between $\avec$ and $\mvec$. It is straightforward to show that this equation determines $\mDc(\Do,\Bo)$ as a decreasing and convex function of $\mDc(\Co,\Ao)$. 
Indeed, the following explicit functional relation between the optimal errors can be proven (\ref{app:cal-bound}); we put $d_a=\mDc(\Co,\Ao)$, $d_b=\mDc(\Do,\Bo)$ for simpler notation:
\begin{equation}\label{eq:cal-bound}
d_a(2-d_a)+d_b(2-d_b)+2\sqrt{d_a(2-d_a)}\sqrt{d_b(2-d_b)}\,\cos\theta=\sin^2\theta.
\end{equation}
This is a closed equation for the lower boundary of the calibration error region, describing the error tradeoff in relation to the incompatibility of $\Ao,
\Bo$. Noting that the left hand side expression, seen as a function $f(d_a,d_b)$, is  increasing  separately in each of its variables as long as $d_a,d_b\le 1$, we obtain the following description of the calibration error region (see also \Fref{fig:calboundary}).
\begin{proposition}\label{prop:cal-mur}
The calibration error region for a pair of maximally incompatible sharp qubit observables $\Ao,\Bo$ is fully determined by a single measurement uncertainty relation, given by the conditional inequality
\begin{equation}\label{eq:cal-mur}
\eqalign{
& d_a(2-d_a)+d_b(2-d_b)+2\sqrt{d_a(2-d_a)}\sqrt{d_b(2-d_b)}\,\cos\theta\ge\sin^2\theta\\
& \qquad\text{if}\ d_a=\mDc(\Co,\Ao)\leq 1,\ d_b=\mDc(\Do,\Bo)\leq 1.
}
\end{equation}
\end{proposition}

Note that in the maximally incompatible case ($\theta=\frac\pi2$), the calibration and metric error regions coincide; their boundary curves are given by the same circular segment as \eref{eq:cal-mur} reduces to the same form as \eref{eq:max}. 

We may now use the relationship  \eref{eq:noise-cal} between measurement noise and calibration error, valid for symmetric approximators, to see the connection between the above relation and an inequality for measurement noise due to Branciard \cite{Branciard2013}. Putting $\varepsilon_a^2=\varepsilon(\Ao,\Co,\rho)^2=2(1-\avec\cdot\cvec)=2d_a$, $\varepsilon_b^2=2d_b$, inequality \eref{eq:cal-mur} turns into Branciard's inequality,
\begin{equation}\label{eq:Branboundqubit}
\eqalign{
	\varepsilon_a^2\left(1-\frac{\varepsilon_a^2}{4} \right) + \varepsilon_b^2\left(1-\frac{\varepsilon_b^2}{4} \right)+ 2\varepsilon_a\varepsilon_b\sqrt{\left(1-\frac{\varepsilon_a^2}{4} \right)\left(1-\frac{\varepsilon_b^2}{4} \right)} \cos\theta \geq \sin^2\theta.
}
\end{equation}
This shows that any experimental qubit test for Branciard's inequality also serves as a test of the calibration error boundary.

\section{Testing measurement uncertainty relations}\label{sec:tests}

We proceed to a brief survey of recent experimental tests of measurement uncertainty relations.  We focus on three methodologies that have been applied and consider the question of which of these qualify as \emph{direct tests} in the sense defined in \cite{BS2015}. 

\subsection{The quest for direct tests}

Generally, a quantum mechanical relation can be tested by experimentally determining the expectation values occurring in it. This may involve performing measurements of different, possibly incompatible, observables. For instance, testing the standard (preparation) uncertainty relation for two incompatible quantities  $\Ao,\Bo$ with first moment operators $A,B$, $\Delta(A,\rho)\Delta(B,\rho)\ge \tfrac{1}{2}\abs{\tr{\rho\comm{A}{B}}}$, requires separate runs of three measurements to obtain the statistics and standard deviations of $\Ao$ and $\Bo$ in a given quantum state, and the expectation value of their commutator. Hence, three (generally) incompatible measurements must be performed. This procedure is in accordance with the interpretation of the given uncertainty relation: one obtains measurement statistics and computes the resulting standard deviations from the data; this enables one to compare these numbers with the theoretical values calculated for the given state. It is understood that the experimenters know which preparation procedures are to be applied to realise the state of interest.

In the case of a measurement uncertainty relation, the task is to compare the theoretical error values and the tradeoff relation they satisfy according to the theory with the result of the error estimation procedure associated with the approximate joint measurement in question. Where this happens, we speak of a direct test of a measurement uncertainty relation. By contrast, in an indirect test, one may take any other suitable measurement that allows one to determine the expectation value(s) involved in the mathematical expression of the error quantity under consideration. We will identify examples of both direct and indirect tests below. We argue that it is only the former that lead to a theory-independent test (see also \cite{BS2015,BLWcoll}).

\subsection{Indirect tests of calibration error tradeoff relations using tests of measurement noise relations}

The first generation of tests of measurement uncertainty relations were concerned with Ozawa's inequality for measurement noise \cite{Erhart2012,Rozema2012}. The values of the noise quantities were determined using the so-called 3-state method \cite{Ozawa2004} and weak measurement method \cite{LW2010}, respectively. These early approaches are analysed in \cite{BLWcoll,BLW2014}, where it is noted that the measurements to be made for obtaining the values of $\varepsilon(\Ao,\Co,\rho)$, $\varepsilon(\Bo,\Do,\rho)$ are not related to error estimation schemes for the approximators $\Co,\Do$. 

This observation applies also to subsequent tests of Branciard's inequality for the noise quantities (e.g., \cite{Ringbauer2014,Kaneda2014}), which use the same methods. Specifically, the three-state method is based on a decomposition of (say) $\varepsilon(\Ao,\Co,\rho)^2$ into a sum of three terms, each of which is an expectation value of a moment of $\Co$ in a different state. No comparison of data for $\Co$ and $\Ao$ is involved.
Similarly, a detailed analysis of the weak-measurement method  given in \cite{BS2015}  shows explicitly that the determination of the so-called weak-valued (hence quasi) probabilities involved in the formula for $\varepsilon(\Ao,\Co,\rho)$ equally fail to reflect any form of error analysis. 

As demonstrated in \cite{BLWcoll}, the noise measure generally fails to be reliable as an error measure if it is applied to approximating observables $\Co$ that do not commute with the target observable $\Ao$. However, the qubit  relations for the noise quantities tested in the experiments cited above show the intuitively expected behaviour of error tradeoff relations. This apparent tension is resolved when we recall that in the qubit case, the measurement noise coincides with the operational calibration error measure (Eq. \eref{eq:noise-cal}) for unbiased binary approximator observables. Accordingly, the above experiments at once constitute indirect tests of the calibration error relation \eref{eq:cal-mur}. 

In revisiting the cited literature, the reader will notice that in some of the papers the focus is on a tradeoff between the error in one measurement and the disturbance this entails on a subsequent measurement. In \cite{BLWcoll} it has been explained in detail that such a sequential execution of two measurements constitutes an instance of a joint measurement, where the disturbance measure for the second component measurement is naturally reinterpreted as an error measure. Hence we have here chosen the language of error tradeoff for joint approximate measurements which encompasses instances of error-disturbance tradeoff relations.

It is interesting to note that \cite{Rozema2012} unwittingly offers a realisation of an optimal approximate joint measurement of maximally incompatible qubit observables $\Ao,\Bo$ (thus with perpendicular $\avec,\bvec$): the joint observable they construct has margins $\Co,\Do$ that are smearings of $\Ao,\Bo$, respectively, such that the calibration errors satisfy the bound \eref{eq:cal-bound}. However, the optimality of this scheme was not known at the time. A simplified and improved scheme for realising this type of joint observable \cite{Kaneda2014} was later used to demonstrate the Branciard bound \eref{eq:Branboundqubit}. A fortiori, this confirms the calibration tradeoff  \eref{eq:cal-mur}. We recall from our earlier analysis that this joint measurement is optimal with respect to both calibration error and metric error.

Incidentally, the highly complicated formula of \cite{LW2010} used to determine the noise quantities via weak measurements becomes independent of the strength parameter when applied to this case of joint observables whose margins $\Co,\Do$ are smearings of $\Ao,\Bo$ and therefore are compatible with their target observables. This goes along with the fact that the so-called weak-valued probabilities appearing in this formula become proper probabilities. This observation can be used to explore what happens if instead of the weak limit, the maximum strength value is chosen. One finds \cite{BS2015} that this turns the (previously weak) measurement scheme preceding the joint measurement into a preparation of the eigenstates of one of the target observables, and one obtains a direct test procedure for measurement noise, and hence for calibration error. Some experimenters (e.g., \cite{Edamatsu2016,SulyokSponar2017}) noticed that the Lund-Wiseman formula for the noise quantities is independent of the strength parameter, and confirmed this experimentally by performing the measurements also in the strong (projective) limit; however, it would be interesting to see whether it is possible to use the data they collected to apply the much simpler, direct method  proposed in \cite{BS2015} to determine the errors, namely, by way of an explicit error analysis.

As is evident from Eq.~\eref{eq:all-cal}, there is an infinity of optimal joint measurements with respect to the calibration error measure if $\avec\perp\bvec$. In \cite{Ringbauer2014}, the case of an intermediate sharp observable $\Mo$ is realised, corresponding to $M=1$ in \eref{eq:all-cal}. Using both the three-state and weak-measurement methods for the determination of the measurement noise values, the bound \eref{eq:Branboundqubit} and hence  \eref{eq:cal-mur} have been confirmed.

\subsection{Direct test of calibration error bound}

A simple realisation of an optimal joint measurement in the form of a sharp intermediate observable $\Mo$ is obtained as follows. Let $\Co$ be a sharp observable,  and let $U$ be a unitary map that transforms the $\pm 1$ eigenstates of $\Co$ into the $\pm 1$ eigenstates of $\Bo$. Then perform the measurement sequence that consists of a projective measurement of $\Co$, followed by the application of $U$ and finally by a projective measurement of $\Bo$.  The associated sequence of non-selective operations (channels) is
\begin{eqnarray*}
\rho&\xrightarrow{\Co}C_+\rho C_++C_-\rho C_-=C_+\tr{\rho C_+}+C_-\tr{\rho C_-}=\rho'\\
&\xrightarrow{U}B_+\tr{\rho C_+}+B_-\tr{\rho C_-}=\tilde\rho\\
&\xrightarrow{\Bo}B_+\tr{\rho C_+}+B_-\tr{\rho C_-}=\tilde\rho.
\end{eqnarray*}
This sequential scheme constitutes a joint observable of a pair $\Co$, $\Do$, with joint probability 
\[
\tr{UC_k\rho C_kU^*B_\ell}=\tr{C_k\rho C_kC_\ell}=\delta_{k\ell}\tr{\rho C_k}.
\] 
It follows that the margins are given by
\[
\tr{\rho D_\pm}=\tr{\tilde\rho B_\pm}=\tr{\rho C_\pm},
\]
hence $\Do=\Co$, and $\Jo_{k\ell}=C_k\delta_{k\ell}$. Note that one could have obtained this equally well by not introducing the 
basis rotation but measuring instead the observable $\Co$ a second time (or not at all). Importantly, this scheme yields a direct
test insofar as it allows an operational procedure for determining the calibration errors from the measured data. To this end one
feeds the device with the eigenstates either of $\Ao$, $\rho=A_\pm$, or of $\Bo$, $\rho=B_\pm$. In either case one finds (noting that $\Do=\Co$):
\[
\eqalign{
2\max\bigl\{|\tr{A_\pm(A_+-C_+)}|,\, |\tr{A_\pm(A_--C_-)}|\bigr\}&=1-\cvec\cdot\avec=\mDc(\Co,\Ao),\\
2 \max\bigl\{|\tr{B_\pm(B_+-D_+)}|,\,|\tr{B_\pm(B_--D_-)}|\bigr\}&=1-\dvec\cdot\bvec=\mDc(\Do,\Bo).
 }
\]

\subsection{Direct test of metric error bound}

The qubit experiment reported in  \cite{XYMZCYFB} tests the various forms of tradeoff relations for metric errors deduced in Section \ref{sec:mer}.  The qubit is realised in terms of two levels of  a trapped ${}^{40}Ca^+$ ion. A joint observable of the form \eref{eq:jt-obs} is chosen such that the compatibility inequalities are saturated, so that $\norm{\cvec\pm\dvec}=1\pm\cvec\cdot\dvec$. This observable has rank-1 effects $J_{k\ell}=\frac14\bigl[(1+k\ell M)I+(k\cvec+\ell\dvec)\cdot\sigvec\bigr]$, with $M=\cvec\cdot\dvec$. A measurement of $\Jo$ is obtained by randomly choosing and measuring one of the sharp observables 
\[
\Lo^{(s)}:k\mapsto \frac12\left[I+ k\frac{\cvec+s\dvec}{1+sM}\right],\quad s=\pm1.
\]
with probabilities $p_s=\frac12(1+sM)$. The joint probability of choosing $\Lo^{(s)}$ and obtaining outcome $k$ is
\[
\eqalign{
p(k,s)&=\tfrac12 (1+sM) \mathrm{tr}\bigl[\rho \Lo^{(s)}(k)\bigr]\\
&=\mathrm{tr}\left[\rho\tfrac12\bigl((1+sM)I+(k\cvec+sk\dvec)\cdot\sigvec\bigr)\right]\\
&=\mathrm{tr}\left[\rho\Jo(k,\ell)\right],\quad (\ell=sk,\ s=k\ell).
}
\]
Measurement of $\Jo$ in this way gives the distributions of $\Co$ and $\Do$ as the margins of $\Jo$. The optimising approximators are implemented by specifying the Bloch vectors $\cvec,\dvec$ according to Eqs.~\eref{eq:BlochDopt} and \eref{eq:YObound}. This is done  for various choices of $\Ao,\Bo$, where $\Ao$ is taken to be the spectral measure of $\sigma_2$ and $\Bo$ that of $\cos\theta\sigma_2+\sin\theta\sigma_3$, for various values of $\theta$ (such as $\theta=\pi/2$, $\pi/4$, $\pi/6$). The distributions of $\Ao$ and $\Bo$ are determined by separate measurements of these sharp target observables. The quantities $\mD(\Co,\Ao)$, $\mD(\Do,\Bo)$ can thus computed directly from the relevant measurement statistics obtained for the states known to yield the worst-case deviations in each case.

These quantities are plotted against the boundary curve of the theoretical error region (\Fref{fig:metric-error-region}), and very good agreement is found. 
There are also plots of the data against the curves given in \eref{eq:D-D-trade} and \eref{eq:sym-opt}, showing an excellent match.

We note that an analogous tight tradeoff relation for error and disturbance, quantified with operational entropic measures, has been 
formulated and tested experimentally in \cite{Sulyok2015}.

\section{Concluding discussion}\label{sec:discuss}

In this paper we revisited the fundamental problem of identifying the ultimate quantum bounds for approximate joint measurements of incompatible observable, exemplified for a pair of sharp qubit observables, such as spin-$\frac12$ components. The problem was posed for the case of the position and momentum of a particle by Heisenberg in his landmark paper of 1927 introducing the uncertainty principle.
He offered an intuitive solution which lacked a rigorous formulation and was of limited generality; but  by restricting himself to approximately preparatory, or repeatable,  measurements he was able to quantify approximation errors (for which no general definition was available at the time) in terms of preparation uncertainty, and thus reduce measurement uncertainty relations to preparation uncertainty relations. 

The connection between preparation uncertainty and measurement error in this special case has led to a conflation of the two ideas. Indeed, one finds prominent textbooks that present the standard (preparation) uncertainty relation  mathematically alongside the correct probabilistic interpretation, but then proceed to paraphrase it in terms of the impossibility of accurate joint measurements of noncommuting quantities or the necessary disturbance of one variable by the measurement of another (e.g., \cite{BohmD,BohmA,Schiff}).

There is, of course, some intuitive plausibility in the idea that one cannot simultaneously \emph{measure} some quantities any better than one can \emph{prepare} them. However, it was only after the advent of a theory of approximate joint measurements, as reviewed here, that possible links between preparation and measurement uncertainty could be investigated rigorously. In the case of  position and momentum it was found that the preparation uncertainty region and error region coincide exactly if described in terms of appropriately matched choices of uncertainty and error measures  \cite{BLW2014b}. This exact correspondence cannot be expected for all pairs of observables, but it is known to hold in the case of systems with a phase space structure and associated translation symmetry \cite{Werner2016}; this includes the case of the phase spaces $\mathbb{Z}_n\times\mathbb {Z}_n$ and hence, in particular, a pair of qubit observables like $\sigma_x,\sigma_z$. Here, in Theorem \ref{thm:pum} and Corollary \ref{cor:MUR=PUR}, using the probabilistic distance as an error measure and an adapted uncertainty measure, we established the coincidence of the uncertainty region and the nontrivial part of the error regions for pairs of maximally incompatible binary qubit observables (with perpendicular Bloch vectors).

The formalisation of the concept of approximate joint measurability is based on the idea of performing a joint measurement of a pair of compatible observables, either of which is taken to represent an approximation of one of the two target observables. For optimal approximations, we saw that the approximating pair of observables need not commute. Their  joint measurability is then accounted for by their being sufficiently \emph{unsharp}. With inequality \eqref{compat-u2} we provided an example of a novel form of uncertainty relation, distinct from both preparation and measurement uncertainty: a tradeoff between the degrees of unsharpness, to be obeyed by two observables in order to ensure their joint measurability. Where the approximating observables are ``smearings" of the target observables (and hence commute with them), the unsharpness is at once a measure of the approximation error. But we found that optimal approximations may be realised by a compatible pair of observables that do not commute with their target observables.

We carried out a comprehensive examination of two quantities that have recently been proposed as measures of error in quantum mechanics, namely the metric error $\mD$ and the calibration error $\mDc$. We described the error regions for both measures, that is, the optimal error bounds for joint approximate measurements of two incompatible binary target observables $\Ao,\Bo$ of a qubit system. We also characterised the optimising compatible pairs $\Co,\Do$ in each case, by identifying their associated Bloch vectors. Finally, we reviewed implementations of experimental tests for associated measurement uncertainty relations and considered which of these qualify as direct tests.

The minimum error curves for approximating $\Ao$ and $\Bo$ by compatible $\Co$ and $\Do$, respectively, were found by solving the corresponding constrained optimisation problems. In the case of the metric error, this was previously done by Yu and Oh \cite{YuOh2014}; here we have added some minor technical clarification and provided a physical interpretation of the associated tradeoff relation as an interplay between the errors, the unsharpness of the approximating observables, and the degree of incompatibility of the target observables. We reviewed an elementary proof for the simple special case of a tangential straight line bound to the metric error region.

On attempting this optimisation problem for the calibration error, by using the method of Lagrange multipliers, we find that a local solution is only possible in the case where the target observables $\Ao,\Bo$ have perpendicular Bloch vectors, that is, when they are maximally incompatible. A simple geometric argument shows that the optimising observables for the calibration error are the extremal (sharp) observables $\Mo$ whose Bloch vectors lie on the surface of the circle spanned by the Bloch vectors of $\Ao$ and $\Bo$, regardless of the degree of incompatibility between the target observables. 

However, in the case of maximally incompatible target observables ($\theta=\pi/2$), there is a continuous family $(\Co_M,\Do_M)$ of compatible pairs of optimising observables, the extremal ones included as $\Mo=\Co_1=\Do_1$ (see \Fref{fig:perp}). In this case the unique  pair of optimal approximating observables found for the metric error is also optimal with respect to the calibration error; is is given by  the pair $(\Co_0,\Do_0)$. 

There is a fundamental difference at this point: The errors $\mD(\Co_M,\Ao)$, $\mD(\Do_M,\Bo)$ increase monotonically with $|M|\in[0,1]$, reflecting the increasing dissimilarity of the observables $\Co_M,\Ao$ and $\Do_M,\Bo$, seen in their increasingly divergent statistics. In contrast the calibration error values, $\mDc(\Co_M,\Ao)$, $\mDc(\Do_M,\Bo)$, are constant. This constancy is related to the fact that the calibration error is less stringent, being a worst-case deviation of distribution pairs only across eigenstates of the target observable  rather than across all states. We note that the definition of $\mDc$ is asymmetric in its arguments in that it requires the reference observable $\Ao$ to be sharp. It is only when both $\Ao,\Co$ are restricted to the set of sharp binary observables that  $\mDc$ becomes a metric. 

The fortuitous coincidence of the calibration error with the noise measure on the set of unbiased binary approximating observables for qubits explains why Branciard's inequality constitutes an error tradeoff with a sound operational interpretation: we rederived this inequality, expressed in terms of calibration errors, as the solution of an optimisation problem, thus exhibiting the lower boundary of the error region.

\ack{Thanks go to Pieter Kok and Mark Pearce for many helpful discussions on the subject of this paper, as well as Oliver Reardon-Smith, Roger Colbeck, Teiko Heinosaari and Stefan Weigert for constructive comments. This work grew out of TB's doctoral research. Support through the White Rose Studentship Network \emph{Optimising Quantum Processes and Quantum Devices for Future Digital Economy Applications} (2011-2014) and the Leverhulme Trust's \emph{Study Abroad Studentship} (2016-2017) is gratefully acknowledged. The final phase of this work was carried out during PB's tenure of a Royal Society Leverhulme Trust Senior Research Fellowship.}

\appendix

\section{}\label{app:necess}
\textbf{Necessity of the compatibility condition \eref{compat}}

We show that the necessity of inequality \eref{compat} follows for the compatibility of a general pair of binary qubit observables $\Co,\Do$, with $C_+=\frac12(c_0 I+\cvec\cdot\sigvec)$,
$D_+=\frac12(d_0 I+\dvec\cdot\sigvec)$ (while sufficiency only holds in the symmetric case, $c_0=d_0=1$); cf. \cite{BH2008}.

The joint measurability conditions \eref{eq:compat-def} applied to two qubit
observables $\Co,\Do$ takes the following form: there exists an
operator $J_{++}=\half(g_0 I+\gvec\cdot\sigvec)$ such that
\begin{equation*}
\eqalign{
J_{++}&\ge 0,\\
J_{+-}&=C_+-J_{++}\ge 0,\\
J_{-+}&=D_+-J_{++}\ge 0,\\ 
J_{--}&=I+J_{++}-C_+-D_+\ge 0.
}
\end{equation*}
Expressing the positivity of these operators in terms of the positivity of their eigenvalues, this set of inequalities is seen to be equivalent to the following:
\begin{equation*}
\eqalign{
\norm{\gvec}&\le g_0;\\
\norm{\cvec-\gvec}&\le c_0-g_0;\\
\norm{\dvec-\gvec}&\le d_0-g_0;\\
\norm{\cvec+\dvec-\gvec}&\le2+g_0-c_0-d_0.
}
\end{equation*}
Let $B(\xvec,r)$ denote the closed ball with centre $\xvec$ and radius
$r$. Then it is seen that the joint measurability of $\Co,\Do$ is
equivalent to the statement that there exists a number $g_0\ge 0$ such
that the intersection of four balls is non-empty:
\begin{equation}\label{eqn:ball-coex}
B(\nulvec,g_0)\,\cap\, B(\cvec,c_0-g_0)\,\cap\,B(\dvec,d_0-g_0)\,\cap\,B(\cvec+\dvec,2+g_0-c_0-d_0)\ne\emptyset.
\end{equation}

The criterion (\ref{eqn:ball-coex}) immediately gives the
following as a necessary condition for joint measurability: the two pairs of balls diagonally opposite to each
other must have separations which are no greater than the sum of
their radii; thus, there must be a $g_0\ge 0$ such that
\begin{equation*}
\eqalign{
\norm{\cvec-\dvec}&\le c_0+d_0-2g_0,\\
\norm{\cvec+\dvec}&\le2-c_0-d_0+2g_0,
}
\end{equation*}
or equivalently,
\begin{equation}\label{eqn:gamma12}
g_1:=\half\norm{\cvec+\dvec}+\half[c_0+d_0-2]\le
g_0\le\half[c_0+d_0]-\half\norm{\cvec-\dvec}=:g_2.
\end{equation}
This gives an interval for $g_0$ to lie in which has to be
nonempty. Therefore the following is a necessary joint measurability
condition:
\begin{equation*}
g_2-g_1=1-[\half\norm{\cvec+\dvec}+\half\norm{\cvec-\dvec}]\ge0.
\end{equation*}

This completes the necessity proof. Returning to the symmetric case, $c_0=d_0=1$, the compatibility conditions reduce to
\begin{equation}\label{eq:sym-com}
B(\nulvec,g_0)\,\cap\, B(\cvec+\dvec,g_0)\,\cap\,B(\cvec,1-g_0)\,\cap\,B(\dvec,1-g_0)\ne\emptyset.
\end{equation}
Now, assuming that $\norm{\cvec+\dvec}+\norm{\cvec-\dvec}\le 2$ holds, and noting that this is equivalent to both of the following inequalities,
\begin{equation}\label{eq:comp-geom}
\tfrac12(1+\cvec\cdot\dvec)\ge \tfrac12\norm{\cvec+\dvec},\quad \tfrac12(1-\cvec\cdot\dvec)\ge \tfrac12\norm{\cvec-\dvec},
\end{equation}
then the choice $g_0=\frac12(1+\cvec\cdot\dvec)$ (which also gives $1-g_0=\frac12(1-\cvec\cdot\dvec)$) satisfies $g_0\ge\frac12\norm{\cvec+\dvec}$
and $1-g_0\ge\frac12\norm{\cvec-\dvec}$. Therefore we may choose $\gvec=\frac12(\cvec+\dvec)$ to see that all inequalities of \eref{eq:sym-com}
are satisfied. Hence,  the choices $g_0=\frac12(1+\cvec\cdot\dvec)$, $\gvec=\frac12(\cvec+\dvec)$ gives rise to a joint observable $\Jo$ of the form \eref{eq:jt-obs},
reconfirming the sufficiency of the compatibility inequality \eref{compat}. 

\begin{figure}[H]
\centering
	\includegraphics[width=0.6\textwidth]{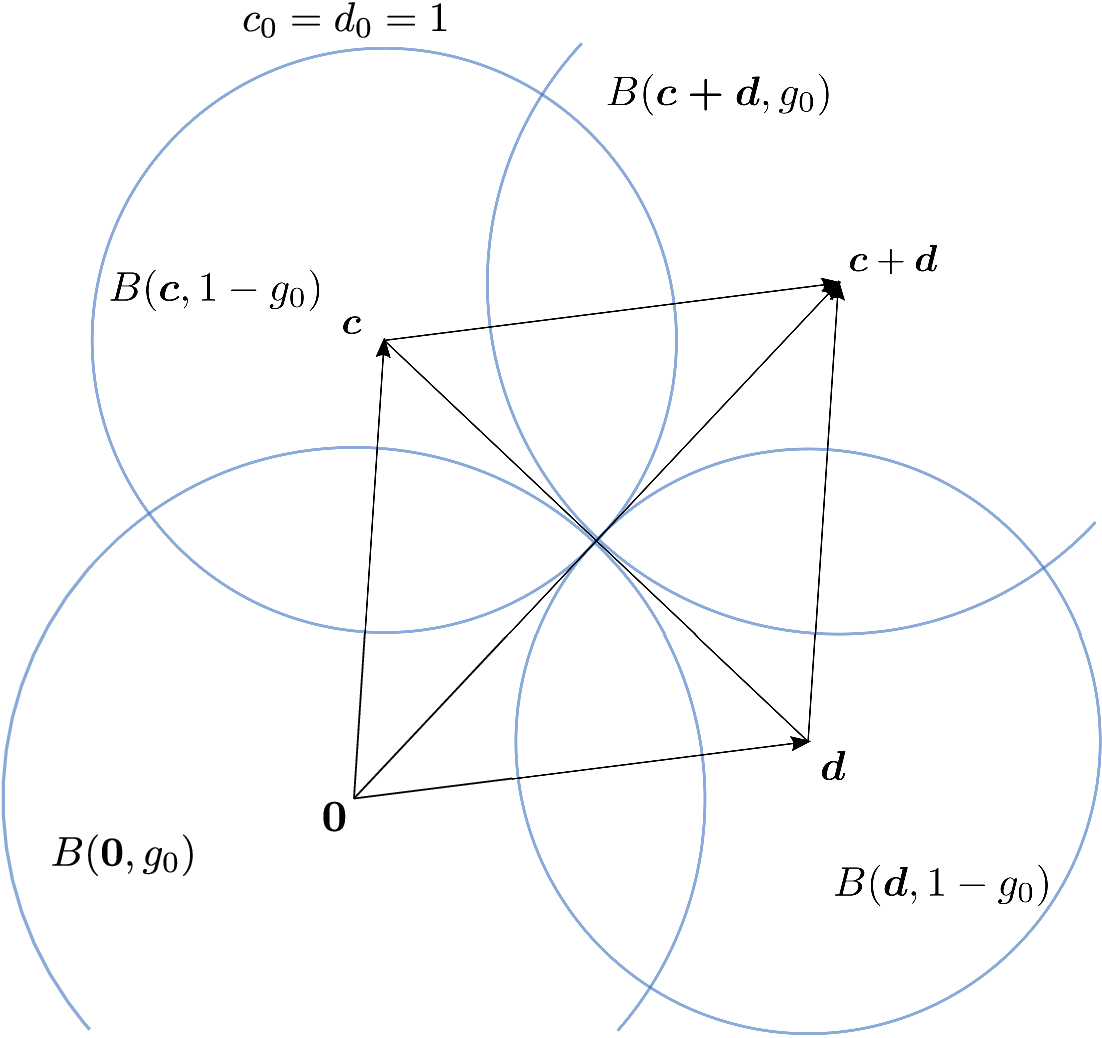}	
	\caption{\label{fig:parallelogram} Illustration of the geometric derivation of \eref{eq:comp-geom}.}
\end{figure}

To summarise, we observe that the compatibility condition \eref{eq:comp-geom} is satisfied if and only if it is possible for the two pairs of diagonally opposite circles  (\Fref{fig:parallelogram}) to be large enough so as to contain the point $\gvec=\frac12(\cvec+\dvec)$.

\section{}\label{app:cal-bound}
\textbf{Derivation of the optimal calibration error relation \eref{eq:cal-bound}}

Here we provide a proof of \eref{eq:cal-bound} based on the fact that optimal error pairs $(d_a,d_b)=\bigl(\mDc(\Co,\Ao),\mDc(\Do,\Bo)\bigr)$ are realised by 
sharp approximating observables, $\Co=\Do$, with $\cvec=\dvec=\mvec$.

\begin{figure}[H]
\centering
	\includegraphics[width=0.6\textwidth]{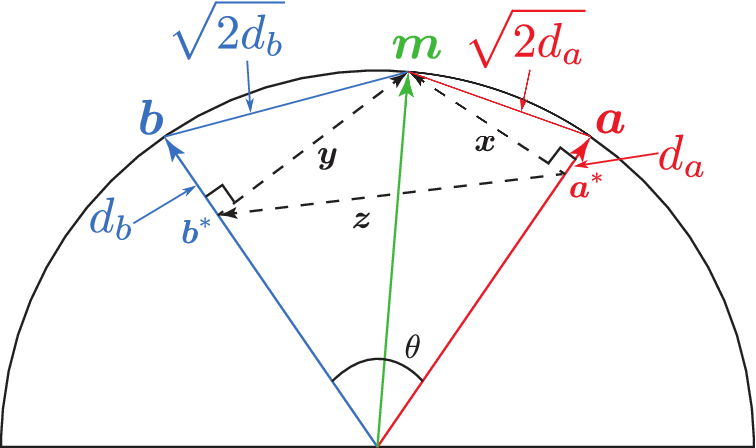}	
	\caption{\label{fig:calbound} Illustrating the derivation of \eref{eq:cal-mur}.}
\end{figure}

The outline for our proof is given in \Fref{fig:calbound}. We begin by choosing a normalised vector $\mvec$ that, due to the argument in Subsection \ref{sec:cal}, we know minimises the calibration error when approximating $\avec$ and $\bvec$. We now rewrite $\mvec$ in the following way:
\begin{equation}\label{eq:maltform}
\eqalign{
	\mvec&= \avec^* + \xvec=\bvec^*+\yvec,
}
\end{equation}
where $\avec^*= (\avec\cdot\mvec)\,\avec$ and $\bvec^*=(\bvec\cdot\mvec)\,\bvec$. By construction, $\xvec\perp\avec$ and $\yvec\perp\bvec$, and from \eref{eq:maltform} we can define the vector $\zvec=\bvec^*-\avec^*=\xvec-\yvec$, so that 
$
	\norm{\zvec}^2 = \norm{\xvec}^2+\norm{\yvec}^2-2\,\xvec\cdot\yvec
$.
By rearranging \eref{eq:maltform} we can find the norms of $\xvec$ and $\yvec$:

\[
\eqalign{
	\norm{\xvec}^2&= \norm{\mvec-\avec^*}^2 = 1- (\avec\cdot\mvec)^2 = d_a(2-d_a),\\
	\norm{\yvec}^2&= \norm{\mvec-\bvec^*}^2 = 1- (\bvec\cdot\mvec)^2 = d_b(2-d_b),
}
\]
and noting that the angle between $\xvec,\yvec$ is $\pi-\theta$, we obtain
\[
\eqalign{
	\xvec\cdot\yvec\, =\norm{\xvec}\norm{\yvec}\cos(\pi-\theta)
	=-\sqrt{d_a(2-d_a)}\sqrt{d_b(2-d_b)}\,\cos\theta.
}
\]
Hence,
\begin{equation}\label{eq:boundst1}
\eqalign{
	\norm{\zvec}^2 =\,   d_a(2-d_a)+d_b(2-d_b)+2\sqrt{d_a(2-d_a)}\sqrt{d_b(2-d_b)}\,\cos\theta.
	}
\end{equation}
The proof is complete once we have shown that $\norm{\zvec}^2=\sin^2\theta$.
To see this, we make use of the form $\zvec=\bvec^*-\avec^*$:
\[
\eqalign{
	\norm{\zvec}^2 &= \norm{\bvec^*-\avec^*}^2 \\
	&= (\avec\cdot\mvec)^2+(\bvec\cdot\mvec)^2 - 2 (\avec\cdot\mvec)(\bvec\cdot\mvec)(\avec\cdot\bvec)\\
	&= 1- \big(1-(\avec\cdot\mvec)^2\big)\big(1-(\bvec\cdot\mvec)^2\big) \\
	&\quad+ \big((\avec\cdot\mvec)(\bvec\cdot\mvec)-(\avec\cdot\bvec)\big)^2 - (\avec\cdot\bvec)^2.
}
\]
Recalling that
\[
	(\avec\cdot\mvec)(\bvec\cdot\mvec)-(\avec\cdot\bvec) = (\mvec\times\avec)\cdot(\mvec\times\bvec),
\]
and noting that the vectors $\mvec\times\avec$ and $\mvec\times\bvec$ are collinear, we have
\[
\eqalign{
	\big((\mvec\times\avec)\cdot(\mvec\times\bvec)\big)^2&=\norm{\mvec\times\avec}^2\norm{\mvec\times\bvec}^2
	=\big(1-(\avec\cdot\mvec)^2\big)\big(1-(\bvec\cdot\mvec)^2\big),
}
\]
and therefore
\[
	\norm{\zvec}^2= 1- (\avec\cdot\bvec)^2 = \norm{\avec\times\bvec}^2=\sin^2\theta.
\]


\section*{References}
\begin{thebibliography}{10}

\bibitem{H27}
Heisenberg,~W.,
\newblock{\"Uber den anschaulichen Inhalt der quantentheoretischen Kinematik und Mechanik,}
\newblock{\em Z. Physik} {\bf 43}, 172 (1927).

\bibitem{QMMT}
Busch, P., Lahti, P., Pellonp\"a\"a, J.-P. and Ylinen, K.,
\newblock{\em Quantum Measurement}, 
\newblock{Springer International Publishing, Switzerland, 2016.}


\bibitem{BHL2007} 
Busch, P., Heinonen (now Heinosaari), T. and Lahti, P.,
\newblock{Heisenberg's Uncertainty Principle},
\newblock{\em Phys. Rep.} \textbf{452}, 155--176 (2007).

\bibitem{BLWcoll}
Busch, P., Lahti,~P. and Werner,~R.F., 
\newblock{Quantum root-mean-square error and measurement uncertainty relations,}
\newblock{\em Rev. Mod. Phys.} \textbf{86}, 1261 (2014).

\bibitem{Ozawa2004}
Ozawa, M.,
\newblock{Uncertainty relations for noise and disturbance in generalized quantum measurements,}
\newblock{\em Ann. Phys. (N.Y.)} {\bf 311}, 350 (2004).

\bibitem{BH2008}
Busch,~P.  and Heinosaari,~T., 
\newblock{Approximate joint measurements of qubit observables}, 
\newblock{\em Quant. Inf. Comp.} \textbf{8}, 0797 (2008).



\bibitem{Branciard2013}
Branciard,~C., 
\newblock{Error-tradeoff and error-disturbance relations for incompatible quantum measurements}, 
\newblock{\em Proc. Nat. Acad. Sci.} \textbf{110}, 6742 (2013).

\bibitem{YuOh2014}
Yu,~S. and Oh,~C.~H., 
\newblock{Optimal joint measurement of two observables of a qubit}, 
\newblock{\em arXiv:} 1402.3785 (2014).

\bibitem{BLW2014}
Busch,~P., Lahti,~P. and Werner,~R.~F., 
\newblock{Heisenberg uncertainty for qubit measurements}, 
\newblock{\em Phys. Rev. A} \textbf{89}, 012129 (2014).

\bibitem{Busch86}
Busch,~P., 
\newblock{Unsharp reality and joint measurements for spin observables}, 
\newblock {\em Phys. Rev. D} \textbf{33}, 2253 (1986).

\bibitem{Busch2009}
Busch, P,
\newblock{On the Sharpness and Bias of Quantum Effects,}
\newblock{\em Found.~Phys.}  \textbf{39}, 712-730 (2009).

\bibitem{BHSS2013}
Busch, P., Heinosaari, T., Schultz, J. and Stevens, N.,
\newblock{Comparing the degrees of incompatibility inherent in probabilistic physical theories,}
\newblock{\em EPL} \textbf{103}, 10002 (2013).


\bibitem{StReHe2008}
Stano, P., Reitzner, D. and Heinosaari, T.,
\newblock{Coexistence of qubit effects,}
\newblock{\em Phys.~Rev.~A} \textbf{78}, 012315 (2008).

\bibitem{BuS2010}
Busch, P. and Schmidt, H.-J.,
\newblock{Coexistence of qubit effects,}
\newblock{\em Quantum Inf.~Process.} \textbf{9}, 143--169 (2010).

\bibitem{Yu2010}
Yu, S., Liu, N.-L., Li, L. and Oh, C.H.,
\newblock{Joint measurement of two unsharp observables of a qubit,}
\newblock{\em Phys.~Rev.~A} \textbf{81}, 062116 (2010).



\bibitem{Miyadera2008}
T.~Miyadera,~T. and Imai,~H.,
\newblock{Heisenberg's uncertainty principle for simultaneous measurement
of positive-operator-valued measures,}
\newblock{\em Phys. Rev. A} {\bf 78}, 052119 (2008).

\bibitem{RSH2017}
Renes, J.M., Scholz, V.B. and Huber, S.,
\newblock{Uncertainty relations: An operational approach to the error-disturbance tradeoff}
\newblock{\em Quantum} {\bf 1} 20 (2017).

\bibitem{BuPe2007}
Busch,~P. and Pearson,~D.B.,
\newblock{Universal joint-measurement uncertainty relation for error bars,}
\newblock{\em J. Math. Phys.} {\bf 48}, 082103  (2007).

\bibitem{Miyadera2011}
Miyadera,~T.,
\newblock{Uncertainty relations for joint localizability and joint measurability in finite-dimensional systems,}
\newblock{\em J. Math. Phys.} {\bf 52}, 072105 (2011).

\bibitem{Buscemi2014} 
Buscemi, F., Hall, M.J.W., Ozawa, M. and Wilde, M.M,
\newblock{Noise and Disturbance in Quantum Measurements: An Information-Theoretic Approach,}
\newblock{\em Phys. Rev. Lett.} {\bf 112}, 050401 (2014).

\bibitem{Ozawa2002}
Ozawa,~M., 
\newblock{Position measuring interactions and the Heisenberg uncertainty principle}, 
\newblock{\em Phys. Lett. A} \textbf{299}, 1 (2002).

\bibitem{Ozawa2013} Ozawa, M.,
\newblock{Disproving Heisenberg's error-disturbance relation,}
\newblock{\em arXiv:}1308.3540v1 (2013).

\bibitem{SulyokSponar2017}
Sulyok,~G. and Sponar,~S., 
\newblock{Is Heisenberg's error-disturbance uncertainty relation violated?: Experimental study of competing approaches}, 
\newblock{\em Phys. Rev. A} \textbf{96}, 022137 (2017).

\bibitem{Bullock2015} Bullock,~T.,
\newblock{From Incompatibility to Optimal Joint Measurability in Quantum Mechanics},
\newblock{PhD Thesis, University of York, 2015}, Available at \href{http://etheses.whiterose.ac.uk/id/eprint/11952}{etheses.whiterose.ac.uk/id/eprint/11952}.

\bibitem{Werner2016} Werner, R.F.,
\newblock{Uncertainty Relations for General Phase Spaces,}
\newblock{\em arXiv:}1601.03843v1 (2016).

\bibitem{BS2015}
Busch,~P. and Stevens,~N., 
\newblock{Direct Tests of Measurement Uncertainty Relations: What It Takes}, 
\newblock{\em Phys. Rev. Lett.} \textbf{114}, 070402 (2015).

\bibitem{Erhart2012}
Erhart, J.,  Sponar, S., Sulyok, G., Badurek, G., Ozawa, M. and Hasegawa, Y.,
\newblock{Experimental demonstration of a universally valid error-disturbance uncertainty relation in spin measurements,}
\newblock{\em Nature Phys.} {\bf 8}, 185 (2012).

\bibitem{Rozema2012}
Rozema,~L.~A., Darabi,~A., Mahler,~D.~H., Hayat,~A., Soudagar,~Y. and Steinberg,~A.~M., 
\newblock{Violation of {H}eisenberg's Measurement-Disturbance Relationship by Weak Measurements}, 
\newblock{\em Phys. Rev. Lett.} \textbf{109}, 100404 (2012).

\bibitem{LW2010}
Lund,~A. and Wiseman,~H.~M., 
\newblock{Measuring measurement-disturbance relationship with weak values}, 
\newblock{\em New J. Phys.} \textbf{12}, 093011 (2010).

\bibitem{Ringbauer2014}
Ringbauer,~M., Biggerstaff,~D.~N., Broome,~M.~A., Fedrizzi,~A., Branciard,~C. and White,~A.~G., 
\newblock{Experimental Joint Quantum Measurements with Minimum Uncertainty}, 
\newblock{\em Phys. Rev. Lett.} \textbf{112}, 020401 (2014).

\bibitem{Kaneda2014}
Kaneda, F. Baek, S.-Y., Ozawa, M. and Edamatsu, K.,
\newblock{Experimental Test of Error-Disturbance Uncertainty Relations by Weak Measurement,}
\newblock{\em Phys. Rev. Lett.} {\bf 112}, 020402 (2014). 

\bibitem{Edamatsu2016} Edamatsu, K.,
\newblock{Quantum measurement and uncertainty relations in photon polarization,}
\newblock{\em Phys. Scr.} {\bf 91}, 073001 (2016).

\bibitem{XYMZCYFB}
Xiong,~T.~P., Yan,~L.~L., Ma,~Z.~H., Zhou,~F., Chen,~L., Yang,~W.~L., Feng,~M. and Busch,~P., 
\newblock{Optimal joint measurements of complementary observables by a single trapped ion}, 
\newblock{\em New J. Phys.} \textbf{19}, 063032 (2017).

\bibitem{Sulyok2015}
Sulyok, G., Sponar, S., Demirel, B., Buscemi, F., Hall, M.J.W., Ozawa, M. and Hasegawa, Y.,
\newblock{Experimental Test of Entropic Noise-Disturbance Uncertainty Relations for Spin-1/2 Measurements,}
\newblock{\em Phys. Rev. Lett.} {\bf 115}, 030401 (2015).


\bibitem{BohmD} Bohm, D.,
\newblock{\emph{Quantum Theory},} 
\newblock{Prentice Hall, New Jersey, 1951.}


\bibitem{BohmA} Bohm, A.,
\newblock{\emph{Quantum Mechanics: Foundations and Applications},} 
\newblock{Springer, New York, 3rd ed. 1993.}


\bibitem{Schiff} Schiff, L.I.,
\newblock{\emph{Quantum Mechanics},} 
\newblock{McGraw-Hill, New York, 3rd ed. 1968.}


\bibitem{BLW2014b}
Busch,~P., Lahti,~P. and Werner,~R.F., 
\newblock{Measurement uncertainty relations,}
\newblock{\em J. Math. Phys.} {\bf 55},  042111 (2014).




\end{thebibliography}
\end{document}